\documentclass[12pt]{article}
\usepackage{theorem}
\theorembodyfont{\rmfamily}
\newtheorem{remark}{Remark}[section]

\title{\textsf{Correspondence between conformal field theory \\and 
Calogero-Sutherland model}}
\author{Reiho Sakamoto$^a$, Jun'ichi Shiraishi$^b$, \\
Daniel Arnaudon$^c$, Luc Frappat$^{cd}$ and Eric Ragoucy$^c$\\
\normalsize{$^a$ \textit{Department of Physics, Graduate School of
Science}} \vspace{-2mm}\\
\normalsize{\textit{University of Tokyo, Hongo, Bunkyo-ku, Tokyo, 113-0033,
Japan}} \\
\normalsize{$^b$ \textit{Graduate School of Mathematical Science, }}
\vspace{-2mm}\\
\normalsize{\textit{University of Tokyo, Komaba, Meguro-ku, Tokyo,
153-8914, Japan}}\\
\normalsize{$^c$ \textit{Laboratoire d'Annecy-le-Vieux de Physique
Th{\'e}orique}} \vspace{-2mm}\\
\normalsize{\textit{LAPTH, CNRS, UMR 5108, Universit{\'e} de Savoie}}
\vspace{-2mm}\\
\normalsize{\textit{B.P. 110, F-74941 Annecy-le-Vieux Cedex, France}} \\
\normalsize{$^d$ \textit{Member of Institut Universitaire de France}} }
\date{July 30, 2004}
\begin{document}
\maketitle

\begin{abstract}
We use the Jack symmetric functions as a basis of the Fock space, and study
the action of the Virasoro generators $L_n$. We calculate explicitly the
matrix elements of $L_n$ with respect to the Jack-basis. A combinatorial
procedure which produces these matrix elements is conjectured. As a
limiting case of the formula, we obtain a Pieri-type formula which
represents a product of a power sum and a Jack symmetric function as a sum of
Jack symmetric functions. Also, a similar expansion was found for the case
when we differentiate the Jack symmetric functions with respect to power
sums. As an application of our Jack-basis representation, a new diagrammatic
interpretation is presented, why the singular vectors of the Virasoro
algebra are proportional to the Jack symmetric functions with rectangular
diagrams. We also propose a natural normalization of the singular vectors
in the Verma module, and determine the coefficients which appear after
bosonization in front of the Jack symmetric functions.
\end{abstract}

\section{Introduction and summary}

In this paper, we revisit the relationship between two theories of quantum
integrable systems, that is, the Virasoro algebra and the
Calogero-Sutherland model.

The Virasoro algebra generated by $L_n\,(n\in \mathbf{Z})$ and the center $c$ with
the commutation relations
\begin{eqnarray}
  \label{eq:Vira}
  [L_n,L_m ]&=&(n-m)\, L_{n+m}+\frac{n(n^2 -1)}{12}\, c\,\delta_{n+m,0},
\end{eqnarray}
was first introduced in the context of the string theory in 1970 \cite{Vi}
and played a prominent role in the conformal field theory initiated by
Belavin, Polyakov and Zamolodchikov in 1984 \cite{BPZ}. If we bosonize the
Virasoro algebra by the Feigin-Fuchs representation, we regard the Virasoro
generators $L_n$'s as operators acting on the bosonic Fock space. This
bosonization technique has been proved to be an extremely powerful tool to
investigate the Virasoro algebra in many situations. This is mainly because
we are able to construct many intertwining operators in terms of the
primary fields and the screening charges. This is in some sense because we
have a good control of the singular or co-singular vectors in the bosonic
Fock space. However, we still do not fully understand the precise
structures of the singular vectors of the Virasoro algebra constructed in
the Verma module. Namely, explicit formulas for the singular vectors are
not very well investigated before bosonization. Therefore, it is desired to
construct a method which generates the Virasoro singular vectors in some
systematic manner.

The Calogero-Sutherland model (CS model) also has its own background. In
1969, Calogero introduced a dynamical system with inverse square repulsive
potential $U=1/ r^2$ and revealed that the model is exactly solvable
\cite{Ca,Ja,Mo}. In 1972, Sutherland extended this model to a periodic
potential case and obtained several exact results (\cite{Su1,Su2}, see also
\cite{CMR}). In 1990's, Ujino, Hikami and Wadati \cite{UHW,UWH}
investigated the complete quantum integrability of these systems. The
Calogero-Sutherland Hamiltonian is
\begin{equation}
  \label{eq:HCS}
  H_{CS}=\sum_{i=1}^{N} \frac{1}{2} p_i^2 +\beta (\beta -1)\sum_{i<j}
  \frac{(\frac{\pi}{L})^2} {\sin^2\frac{\pi}{L}(q_i -q_j )},
\end{equation}
where $q_i\, (0\leq q_i \leq L)$ are the coordinates and
$p_i =-\sqrt{-1}\frac{\partial}{\partial q_i}$ are the momenta.
The Calogero-Sutherland model is a quantum integrable system in the
following sense. First, the exact ground state of $H_{CS}$ has a simple
factorized form
\begin{equation}
  \psi_{0}=\prod_{i<j} \sin^\beta \frac{\pi}{L}(q_i-q_j),
\end{equation}
and all the excited states can be given by multiplying certain symmetric
polynomials to $\psi_0$ as
\begin{equation}
  \psi_\lambda=P_\lambda(x_1,x_2,\cdots,x_N;\beta) \psi_0,
\end{equation}
where $x_i=\exp({2\pi \sqrt{-1}\over L}q_i)$. Here,
$P_\lambda(x_1,x_2,\cdots,x_N;\beta)$'s are called the Jack symmetric
polynomials \cite{Fo}, and are our central object in this paper \cite{Jack,St,Mac2}. The
Jack symmetric functions have various interesting combinatorial properties,
and of course play an essential role in studies of the Calogero-Sutherland
model (see for example \cite{Ha1,Ha2}).

There is another aspect in the theory of the Jack symmetric polynomials. If
one tries to find a good representation theory of the Virasoro algebra, the
Jack polynomial automatically comes into the game by some mysterious
reasons. The first example of such phenomena was found by Mimachi and
Yamada \cite{TK,MM}. They found that the singular vectors of the Virasoro
algebra represented in the bosonic Fock space are nothing but the Jack
symmetric polynomials with rectangular diagrams. Our main aim in the
present article is to introduce a full-use of the Jack symmetric functions
for the Feigin-Fuchs representation of the Virasoro algebra, and have a
much better understanding of these Virasoro-Jack correspondence. \bigskip

We show our basic idea using some simple examples. Let us introduce a
bosonic operators $a_n\,(n\in \mathbf{Z})$
\begin{equation}
  [a_n,a_m]=n\,\delta_{n+m,0},
\end{equation}
and the Fock space with the Fock vacuum $\langle A| a_0 =A\langle A| $.
Then, we can represent the generators $L_n$'s of the Virasoro algebra by
these boson operators -- the Feigin-Fuchs representation \cite{FF1}
(see Eq.(\ref{FeFu})). Then,
we can use the Jack symmetric functions as a basis of the Fock space. This
mapping is simply given by replacing the power sum functions $p_n$'s by the
bosonic modes $a_n/\sqrt{2\beta}$'s \cite{AMOS1,AMOS2}. See also
\cite{Mac2}. Thus we have bosonization of the Jack symmetric functions, and
we can study the action of $L_n$'s on the Jack symmetric functions.

Deferring the detailed definition to section 2, we give one impressive example.
\begin{eqnarray*}
\langle A_{r+1,s+1}|J
\unitlength 5pt
\begin{picture}(3,2)(-0.2,1.4)
\multiput(0,0)(1,0){3}{\line(0,1){2}}
\put(0,0){\line(1,0){2}}
\multiput(0,1)(0,1){2}{\line(1,0){3}}
\put(3,1){\line(0,1){1}}
\end{picture}\,\,
L_1
&=&
\langle A_{r+1,s+1}|\,\sqrt{2\beta}\left( \,
\frac{\beta\, (1+2\beta)}{(1+\beta)(3+2\beta)}A_{r-1,s-4}
J\unitlength 5pt
\begin{picture}(4,2)(-0.2,1,4)
\multiput(0,0)(1,0){3}{\line(0,1){2}}
\put(0,0){\line(1,0){2}}
\multiput(0,1)(0,1){2}{\line(1,0){4}}
\multiput(3,1)(1,0){2}{\line(0,1){1}}
\end{picture}
\right.\\
& &\qquad\qquad\quad\qquad\,\,\,\,
 +\frac{\beta}{(1+\beta)(2+\beta)}
A_{r-2,s-3} J
\unitlength 5pt
\begin{picture}(4,2)(-0.2,1,4)
\multiput(0,0)(1,0){4}{\line(0,1){2}}
\multiput(0,0)(0,1){3}{\line(1,0){3}}
\end{picture}\\
& &\qquad\qquad\quad\qquad
\left. +\frac{2\, (3+\beta)}{(2+\beta)(3+2\beta)}
A_{r-3,s-1}J
\unitlength 5pt
\begin{picture}(4,2)(-0.2,1,4)
\multiput(0,-1)(1,0){2}{\line(0,1){3}}
\put(0,-1){\line(1,0){1}}
\put(0,0){\line(1,0){2}}
\multiput(0,1)(0,1){2}{\line(1,0){3}}
\put(2,0){\line(0,1){2}}
\put(3,1){\line(0,1){1}}
\end{picture}\,
\right).
\end{eqnarray*}
In this example, we immediately notice that the operation of the $L_1$ has
the effect of adding one box to each possible place of the Young diagram,
and factorized coefficients can be seen in front of each Jack symmetric
functions. We performed careful calculations using Jack symmetric functions
up to degree 10, and we found that the same structure do exist in general.

Firstly, for positive $n$, the effect of $L_n$ is represented by adding
$n$-boxes in each possible ways to the original diagram. On the other hand,
to apply the $L_{-n}$ means to remove $n$ boxes in each possible ways.
The crucial point for general $n$ is that, for applying $L_n$ (or
$L_{-n}$), if we keep track of the ways to add (or subtract) the boxes from
the original Young diagram, we have a good combinatorial rule to calculate
the corresponding coefficients in a nice factorized form. Namely, the
rational functions like ${\beta(1+2\beta)\over (1+\beta)(3+2 \beta)}$ in
the above example can be easily obtained by the combinatorial rule. In other
words, if there are $m$ possible ways to make one particular Young diagram
(say $\mu$) from the original diagram (say $\lambda$), then $m$-terms with
$\langle A|J_\mu$ will appear in $\langle A| J_\lambda L_n$. If we sum up
these $m$-terms, we do not easily see any good combinatorial structure in
the coefficient of $\langle A|J_\mu$. So, our decomposition with respect to
the possible ways is crucial to our aim.

Second, as in the above example, we have a parameter
$A_{r,s}=\frac{1}{\sqrt{2}} \left( r\sqrt{\beta}-s\frac{1}{\sqrt{\beta}}
\right)$ for the Fock vacuum. We can see that the action of $L_n$ depends
on this parameter in a systematic manner. As is easily seen from the
example, we have a coefficient of the shape $A_{r-*,s-*'}$ in each term, and
this shift $(*,*')$ tells us the coordinate to where the box is added. For
example, in the first term in R.H.S. of the above example, $A_{r-1,s-4}$
appears, and added box in this term is at first row, fourth column of the
Young diagram.

\bigskip

As an application of this Jack-basis, we are able to have another way to 
understand the Mimachi-Yamada theorem \cite{MM}. Actually, we can do a little
better in the following sense. If we forget about the rational factors
depending on $\beta$ and keep the factors of the shape $A_{r-*,s-*'}$ for
simplicity, the action of the Virasoro generators looks like
\begin{eqnarray*}
L_1 J\unitlength 5pt
\begin{picture}(4,2)(-0.2,1,4)
\multiput(0,0)(1,0){4}{\line(0,1){2}}
\multiput(0,0)(0,1){3}{\line(1,0){3}}
\end{picture}
|A_{r+1,s+1}\rangle
&=&\,\,\,
A_{r-2,s-3}J\unitlength 5pt
\begin{picture}(4,2)(-0.2,1.4)
\multiput(0,0)(1,0){3}{\line(0,1){2}}
\put(0,0){\line(1,0){2}}
\multiput(0,1)(0,1){2}{\line(1,0){3}}
\put(3,1){\line(0,1){1}}
\end{picture}
|A_{r+1,s+1}\rangle ,\\
L_2 J\unitlength 5pt
\begin{picture}(4,2)(-0.2,1,4)
\multiput(0,0)(1,0){4}{\line(0,1){2}}
\multiput(0,0)(0,1){3}{\line(1,0){3}}
\end{picture}
|A_{r+1,s+1}\rangle
&=&\,\,\,
A_{r-2,s-3}J
\unitlength 5pt
\begin{picture}(4,2)(-0.2,1.4)
\multiput(0,0)(1,0){2}{\line(0,1){2}}
\put(0,0){\line(1,0){1}}
\multiput(0,1)(0,1){2}{\line(1,0){3}}
\multiput(2,1)(1,0){2}{\line(0,1){1}}
\end{picture}
|A_{r+1,s+1}\rangle\\
& &+A_{r-2,s-3}J
\unitlength 5pt
\begin{picture}(3,2)(-0.2,1.4)
\multiput(0,0)(1,0){3}{\line(0,1){2}}
\multiput(0,0)(0,1){3}{\line(1,0){2}}
\end{picture}
|A_{r+1,s+1}\rangle .
\end{eqnarray*}
Note that from the commutation relation (\ref{eq:Vira}), we can
generate all the 
$L_n,n>0$ only from $L_1$ and $L_2$. Note also that $A_{0,0}=0$. Thus we
can conclude that $J \unitlength 5pt 
\begin{picture}(4,2)(-0.2,1,4)
\multiput(0,0)(1,0){4}{\line(0,1){2}} \multiput(0,0)(0,1){3}{\line(1,0){3}}
\end{picture}
|A_{r,s}\rangle$ is a singular vector if $r=2,s=3$, i.e. when the highest
weight is equal to $A_{2+1,3+1}$. Usualy we call this singular vector
$|\chi _{2,3}\rangle$ etc.

At present, we still do not have any good idea how we study the precise
form of the Virasoro singular vectors in the Verma module. However, the
following observation tells us that there may be a natural hidden structure
in the Verma module which corresponds to the Jack-basis after bosonization.
Let us try to normalize the singular vector $|\chi_{r,s}\rangle$ at level
$n=rs$ as
\begin{equation}
  |\chi_{r,s}\rangle=\left(c_1 L_{-n}+c_2 L_{-n+1}L_{-1}+\cdots+ 1\times
  L_{-1}^n\right) |A_{r+1,s+1}\rangle,
\end{equation}
by letting the coefficient of $L_{-1}^n$ being unity. Then, in the lecture
note \cite{Shi}, it was conjectured that after bosonization of the
normalized singular vector, we get
\begin{eqnarray}
  |\chi _{r,s}\rangle &=& \prod_{i=1}^r \prod_{j=1}^s (i\beta -j)\cdot
  J_{(s^r )} |A_{r+1,s+1}\rangle ,\\
  \langle \chi_{r,s}|&=&0 .
\end{eqnarray}
In section 4.2,
we prove these equalities.
\bigskip

Let us consider a limiting case of the matrix elements for $L_n$. If we
take the limit $A_{r,s} \to \infty$, then $L_n$ becomes proportional to
$p_n$ or $\partial/\partial p_n$. Thus in this limit, we obtain a formula
for product of power sum $p_n$ to the Jack symmetric functions, or for
differentiating the Jack symmetric functions with respect to $p_n$. These
formulas can be easily obtained just by dropping all the factors of the
shape $A_{r-*,s-*'}$ from the original formula. For example, in connection
with the above example, we have
\begin{eqnarray*}
J\unitlength 5pt
\begin{picture}(3,2)(-0.2,1.4)
\multiput(0,0)(1,0){3}{\line(0,1){2}}
\put(0,0){\line(1,0){2}}
\multiput(0,1)(0,1){2}{\line(1,0){3}}
\put(3,1){\line(0,1){1}}
\end{picture}\,\,
p_1
&=&
\left( \,
\frac{\beta\, (1+2\beta)}{(1+\beta)(3+2\beta)}
J\unitlength 5pt
\begin{picture}(4,2)(-0.2,1,4)
\multiput(0,0)(1,0){3}{\line(0,1){2}}
\put(0,0){\line(1,0){2}}
\multiput(0,1)(0,1){2}{\line(1,0){4}}
\multiput(3,1)(1,0){2}{\line(0,1){1}}
\end{picture}
\right.\\
& &\,\,\,
 +\frac{\beta}{(1+\beta)(2+\beta)}
J\unitlength 5pt
\begin{picture}(4,2)(-0.2,1,4)
\multiput(0,0)(1,0){4}{\line(0,1){2}}
\multiput(0,0)(0,1){3}{\line(1,0){3}}
\end{picture}\\
& &\,
\left. +\frac{2\, (3+\beta)}{(2+\beta)(3+2\beta)}
J\unitlength 5pt
\begin{picture}(4,2)(-0.2,1,4)
\multiput(0,-1)(1,0){2}{\line(0,1){3}}
\put(0,-1){\line(1,0){1}}
\put(0,0){\line(1,0){2}}
\multiput(0,1)(0,1){2}{\line(1,0){3}}
\put(2,0){\line(0,1){2}}
\put(3,1){\line(0,1){1}}
\end{picture}\,
\right).
\end{eqnarray*}
\\

Let us mention that
in \cite{CDK}, the authors give the singular vectors for the $c <1$ Fock
modules over 
the Virasoro algebra in terms of Schur polynomials. This is done 
explicitly for the cases:  $(r,s) = (1,s),(2,s),(r,1),(r,2),(3,3)$ 
and up to coefficients for the general case (see also the appendix for
the action of the $L_n$).

In \cite{MaSc} a correspondence between Calogero--Sutherland model and
correlators of vertex operators of a CFT with $U(1)$ symmetry was also
established.
\\[3mm]

The plan of our paper is as follows. In section 2, we briefly review how to
bosonize the Virasoro algebra and Jack symmetric functions. In section 3,
we give formulas for action of $L_n$ operators on the Jack symmetric
functions, and describe related formula of symmetric function. In section
4, we give the application of these formulas to the theory of singular
vectors. A natural normalization of the Virasoro singular vectors in the
Verma module is discussed, and whose bosonization is discussed.

\section{Virasoro algebra and Jack symmetric functions}

\subsection{Fock space representation of the Virasoro algebra}

In this section, we recall the Feigin-Fuchs representation of the Virasoro
algebra. Introduce the Fock vacuum $|0\rangle$ by
\begin{eqnarray}
  a_n |0\rangle &=&0\qquad (n \geq 0),\\
  \langle 0|0\rangle&=&1 \, .
\end{eqnarray}
and the zero-mode operator $Q$ by the commutation relation
\begin{equation}
  [a_n ,Q]=\delta _{n,0}.
\end{equation}
Define the eigenstates of the zero-mode $a_0$ by
\begin{equation}
  |A\rangle = e^{A Q}|0\rangle,\quad
  \langle A | = \langle 0 | e^{-A Q},
\end{equation}
satisfying
\begin{equation}
  a_0 |A\rangle = A |A\rangle ,\quad
  \langle A |a_0 = \langle A |A .
\end{equation}
The Fock space $\mathcal{F}_A$ is then defined by
\begin{equation}
  \mathcal{F}_A = \mathbf{C}[a_{-1},a_{-2},a_{-3},\cdots ]|A\rangle .
\end{equation}
We also define the dual space of $\mathcal{F}_A$ as
\begin{equation}
  \mathcal{F}_A^* = \langle A|\mathbf{C}[a_1,a_2,a_3,\cdots ].
\end{equation}

If the central charge of the Virasoro algebra satisfies the condition
\begin{equation}
  c=1-\frac{6(\beta -1)^2}{\beta} ,
\end{equation} 
then we can represent $L_n$'s as (Feigin-Fuchs representation)
\begin{equation}
  L_n = \frac{1}{2} \sum_{k\in \mathbf{Z}} :a_{n-k}a_k :-A_{1,1} (n+1) a_n .
  \label{FeFu}
\end{equation}
Here, we used the usual normal ordering rule
\begin{equation}
  :a_n a_m:\, = \cases{ 
  a_n a_m & (if $n\leq m$), \cr 
  a_m a_n & (if $n>m$), \cr}
\end{equation}
and the notation
\begin{equation}
  A_{r,s}=\frac{1}{\sqrt{2}}\left( r\sqrt{\beta}
  -s\frac{1}{\sqrt{\beta}}\right).
\end{equation}
Note that the parameter $\beta$ is related to the central charge
of the Virasoro algebra as $ c=1-\frac{6\, (\beta -1)^2}{\beta}=1-12A_{1,1}^2$
and this $\beta$ also will be related to the coupling constant for the
Calogero-Sutherand model, or in other words, the Jack symmetric polynomial.

\subsection{Jack symmetric function}

We now turn to the Calogero-Sutherland model and consider the
Schr\"{o}dinger equation defined by the Hamiltonian $H_{CS}$
(\ref{eq:HCS}). Consider 
$N$ particles moving on a circle of length $L$. Let $q_i$'s be the
coordinates of the particles. Then their chord distance is given by
\begin{equation}
  \frac{L}{\pi} \sin \left( \frac{\pi}{L}(q_i -q_j )\right) \, .
\end{equation}
Thus the interaction term in (\ref{eq:HCS}) expresses a strong
repulsive interaction 
between the particle moving on the circle with the potential $1/r^2$. Set
$x_i=\exp ({\frac{2\pi \sqrt{-1}}{L}q_i})$. Then, using the exact ground
state of $H_{CS}$ obtained by Sutherland, we can write all the eigen
functions of the CS model as
\begin{equation}
  \left( \prod_i x_i \right) ^{l-\beta \frac{N-1}{2}} \cdot \prod_{i<j}
  (x_i -x_j )^{\beta} \cdot P_{\lambda}\, ,
\end{equation}
where $l\in \mathbf{Z}$ and $P_{\lambda}$ are some symmetric polynomials in $x_i$'s
parameterized by partitions (or Young diagrams) $\lambda$. Then we notice
that the $P_{\lambda}$'s are the eigenstates of the operator
\begin{equation}
  H_{\beta}:=\sum_{i=1}^{N} \left( x_i \frac{\partial}{\partial x_i}
  \right) ^2 + \beta\sum_{i<j} \frac{x_i +x_j }{x_i -x_j}\left( x_i
  \frac{\partial}{\partial x_i}-x_j \frac{\partial}{\partial x_j} \right)
  \, .
\end{equation}
Namely, the $P_{\lambda}$'s are completely characterized by the
Schr\"{o}dinger equation
\begin{eqnarray}
  H_{\beta}P_{\lambda}(x_1,x_2,\cdots x_N) &=& \epsilon
  _{\lambda}P_{\lambda}(x_1,x_2,\cdots x_N),\\
  \epsilon _{\lambda} &=& \sum_{i=1}^{N} \left( \lambda_i^2 +\beta\,
  (N+1-2i)\,\lambda_i \right) ,
\end{eqnarray} 
and the expansion of the form
\begin{equation}
P_{\lambda}=m_{\lambda}+\sum_{\mu <\lambda}c_{\lambda \mu}(\beta )m_{\mu} . \label{expansion}
\end{equation}
Here, $m_{\lambda}$'s denote the monomial symmetric polynomials.
These $P_{\lambda}$ are called Jack symmetric polynomials \cite{Su2,St}. 

For our purpose, we need to represent the Jack symmetric polynomials in
terms of the power sums $p_n$. To this end, we need to let $N$ to be
infinity. Such symmetric polynomials with infinitely many variables are called
symmetric functions. So, in what follows, we consider Jack symmetric
functions.

The Jack symmetric functions are uniquely characterized 
as a system of orthogonal functions with respect to
the inner product
\begin{equation}
  \langle p_{\lambda},p_{\mu}\rangle :=\delta_{\lambda,\mu}
  z_{\lambda}\beta^{-l(\lambda)}, \label{inner_product}
\end{equation}
if we work with expansion (\ref{expansion}).
Here, the power sum functions $p_n$ are
\begin{equation}
  p_n :=\sum_{i=1}^{\infty} x_i^n ,
\end{equation}
and also write for a partition
$\lambda =(\lambda_1 ,\lambda _2 ,\lambda_3 ,\cdots)$ as
\begin{equation}
  p_{\lambda}=\prod _{i=1}^{l(\lambda )} p_{\lambda _i} ,
\end{equation}
where $l(\lambda)$ is length of $\lambda$, i.e. the largest $i$ for which
$\lambda _i \ne 0$. We also define the weight of $\lambda$ as $|\lambda| =
\sum_{i}\lambda_{i}$. The length $l(\lambda)$ and the weight $|\lambda|$
are equal respectively to the number of rows and the number of boxes of the
Young diagram $\lambda$. We have set for a partition $\lambda=(\cdots
3^{m_3} 2^{m_2} 1^{m_1})$
\begin{equation}
  z_{\lambda} :=\prod_i i^{m_i}\cdot m_i !\, .
\end{equation}
Just from the inner product (\ref{inner_product}), 
we find a relationship between the Jack symmetric functions and the bosonic
Fock space. In fact, we can identify these two by
\begin{eqnarray}
  p_n &\leftrightarrow & \sqrt{\frac{2}{\beta}}\, a_{-n} |A\rangle
  ,\label{liden}\\
  p_n &\leftrightarrow & \langle A|a_n \frac{1}{\sqrt{2\beta\,}}
  .\label{riden}
\end{eqnarray}
More precisely we can show
\begin{equation}
  \left( \langle A|\prod_{i=1}^{l(\lambda)} a_{\lambda_i}
  \frac{1}{\sqrt{2\beta\,}}\right) \left( \prod_{j=1}^{l(\mu)}
  \sqrt{\frac{2}{\beta}}\, a_{-\mu_j}|A\rangle\right) =
  \delta_{\lambda,\mu} z_{\lambda}\beta^{-l(\lambda)}.
\end{equation}
Thus we can use these boson operators to represent the inner product
(\ref{inner_product}), hence from the Macdonald's existence theorem, we can
represent the Jack symmetric functions by boson operators\footnote{See page 379 
of Macdonald's book \cite{Mac2} as to
the meaning of the expansion (\ref{expansion}).}. Note that we
will identify two $\beta$'s which appeared in $H_{CS}$ and in the central
charge of the Virasoro algebra. We will see under this identification,
there is a good relationship between the Virasoro algebra and the
Calogero-Sutherland model.

There are several ways to normalize the Jack symmetric functions. We use
the so-called integral form $J_\lambda$. To describe this normalization, we
need some terminology (see section I.1 of \cite{Mac2}). Define the
coordinate of Young diagram $\lambda = \left(
\lambda_1,\lambda_2,\lambda_3,\cdots \right)$ like as a matrix, i.e. if the box
$s\in\lambda$ is on the $i$-th row and $j$-th column, then the coordinate
of $s$ is $(i,j)$. The conjugate of partition $\lambda$ is a transpose of
$\lambda$ with respect to the main diagonal when expressed by diagram, and
denoted by $\lambda^{'}$. The arm length $a(s)$ and the leg length $l(s)$
of $s=(i,j)\in\lambda$ is defined by
\begin{equation}
  a(s)=\lambda_i -j\, ,\qquad l(s)=\lambda_j^{'} -i\, ,
\end{equation}
where $\lambda_j^{'}$ means the $j$-th element of $\lambda^{'}$. We
linearly extend the definition of $a(s)$ and $l(s)$ when $s$ is outside of
diagram.

In this section, we have been working with $P_{\lambda}$, whose norm is \cite{St}
\begin{equation}
  \langle P_{\lambda},P_{\mu}\rangle = \delta_{\lambda ,\mu} \prod_{s\in
  \lambda}\frac{a(s)+\beta\, l(s)+1}{a(s)+\beta\, l(s)+\beta}\, .
  \label{orthogonality}
\end{equation}
We give degree 3 of $P_\lambda$'s as examples.
\begin{eqnarray*}
P\unitlength 4pt
\begin{picture}(3,0)
\multiput(0,0)(1,0){4}{\line(0,1){1}}
\multiput(0,0)(0,1){2}{\line(1,0){3}}
\end{picture}
&=&
\frac{2}{(1+\beta)(2+\beta)} \; p_3 +\frac{3\beta}{(1+\beta)(2+\beta)}
\; p_2 p_1 +\frac{\beta^2}{(1+\beta)(2+\beta)} \; p_1^3 \\
P \unitlength 4pt
\begin{picture}(3,0)(0,1)
\multiput(0,0)(1,0){2}{\line(0,1){2}}
\put(0,0){\line(1,0){1}}
\multiput(0,1)(0,1){2}{\line(1,0){2}}
\put(2,1){\line(0,1){1}}
\end{picture}
&=&
-\frac{1}{(1+2\beta)} \; p_3+\frac{(1-\beta)}{(1+2\beta)} \; p_2 p_1
+\frac{\beta}{(1+2\beta)} \; p_1^3\\
P \unitlength 4pt
\begin{picture}(3,3)(0,2)
\multiput(0,0)(1,0){2}{\line(0,1){3}}
\multiput(0,0)(0,1){4}{\line(1,0){1}}
\end{picture}
&=&
\frac{1}{3} \; p_3-\frac{1}{2} \; p_2 p_1+\frac{1}{6} \; p_1^3
\end{eqnarray*}

However, it will become clear that
integral form of the Jack symmetric functions $J_\lambda$ 
\begin{equation}
  J_\lambda =\prod_{s\in \lambda}\left( \frac{a(s)}{\beta}+l(s)+1\right)
  P_\lambda\, ,
\end{equation}
are more suitable for our purpose.
Note that, when we use this integral form, the Jack symmetric functions
become the polynomial of $\alpha =1/\beta$ and $p_n$ where all the
coefficients are integer. We also notice that in Jack symmetric function of
degree $n$, the coefficient in front of $p_1^n$ is normalized as unity. We
give a list of Jack symmetric functions up to degree 4.
\begin{eqnarray*}
J\unitlength 4pt
\begin{picture}(4,1)
\multiput(0,0)(1,0){2}{\line(0,1){1}}
\multiput(0,0)(0,1){2}{\line(1,0){1}}
\end{picture}
&=& p_1\\
J\unitlength 4pt
\begin{picture}(4,1)
\multiput(0,0)(1,0){3}{\line(0,1){1}}
\multiput(0,0)(0,1){2}{\line(1,0){2}}
\end{picture}
&=& \alpha p_2+p_1^2\\
J\unitlength 4pt
\begin{picture}(4,2)(0,1)
\multiput(0,0)(1,0){2}{\line(0,1){2}}
\multiput(0,0)(0,1){3}{\line(1,0){1}}
\end{picture}
&=& -p_2+p_1^2\\
J\unitlength 4pt
\begin{picture}(4,1)
\multiput(0,0)(1,0){4}{\line(0,1){1}}
\multiput(0,0)(0,1){2}{\line(1,0){3}}
\end{picture}
&=& 2\alpha^2 p_3+3\alpha p_2 p_1+p_1^3\\
J\unitlength 4pt
\begin{picture}(4,2)(0,1)
\multiput(0,0)(1,0){2}{\line(0,1){2}}
\put(0,0){\line(1,0){1}}
\multiput(0,1)(0,1){2}{\line(1,0){2}}
\put(2,1){\line(0,1){1}}
\end{picture}
&=& -\alpha p_3+(\alpha -1)p_2 p_1+p_1^3\\
J\unitlength 4pt
\begin{picture}(4,3)(0,2)
\multiput(0,0)(1,0){2}{\line(0,1){3}}
\multiput(0,0)(0,1){4}{\line(1,0){1}}
\end{picture}
&=& 2p_3-3p_2 p_1+p_1^3\\
J\unitlength 4pt
\begin{picture}(4,1)
\multiput(0,0)(1,0){5}{\line(0,1){1}}
\multiput(0,0)(0,1){2}{\line(1,0){4}}
\end{picture}
&=& 6\alpha^3 p_4 +8\alpha^2 p_3 p_1 +3\alpha^2 p_2^2+6\alpha p_2 p_1^2
+p_1^4\\
J\unitlength 4pt
\begin{picture}(4,2)(0,2)
\multiput(0,0)(1,0){2}{\line(0,1){2}}
\put(0,0){\line(1,0){1}}
\multiput(0,1)(0,1){2}{\line(1,0){3}}
\multiput(2,1)(1,0){2}{\line(0,1){1}}
\end{picture}
&=& -2\alpha^2 p_4+2\alpha (\alpha -1)p_3 p_1-\alpha p_2^2 +(3\alpha -1)p_2
p_1^2 +p_1^4\\
J \unitlength 4pt
\begin{picture}(4,2)(0,1)
\multiput(0,0)(1,0){3}{\line(0,1){2}}
\multiput(0,0)(0,1){3}{\line(1,0){2}}
\end{picture}
&=& -\alpha (\alpha -1)p_4-4\alpha p_3 p_1+(1+\alpha +\alpha^2)p_2^2 +
2(\alpha -1)p_2 p_1^2 +p_1^4\\
J \unitlength 4pt
\begin{picture}(4,3)(0,2)
\multiput(0,0)(1,0){2}{\line(0,1){3}}
\multiput(0,0)(0,1){2}{\line(1,0){1}}
\multiput(0,2)(0,1){2}{\line(1,0){2}}
\put(2,2){\line(0,1){1}}
\end{picture}
&=& 2\alpha p_4-2(\alpha -1)p_3 p_1-\alpha p_2^2+(\alpha -3)p_2 p_1^2 +
p_1^4\\
J \unitlength 4pt
\begin{picture}(4,4)(0,3)
\multiput(0,0)(1,0){2}{\line(0,1){4}}
\multiput(0,0)(0,1){5}{\line(1,0){1}}
\end{picture}
&=& -6p_4 +8p_3 p_1+3p_2^2 -6p_2 p_1^2 +p_1^4
\end{eqnarray*}
\bigskip

The general form of the Jack symmetric functions $J_{\lambda}$ is
\begin{equation}
  J_{\lambda} = \sum_{|\mu|=|\lambda|} c_{\lambda\mu} \, p_{\mu}
\end{equation}
with the normalization given by $c_{\lambda\mu}=1$ for
$\mu=(1^{|\lambda|})$.

\section{Actions of Virasoro operators on Jack symmetric functions}
In this section, we consider $\beta$, $r$ and $s$ as general real numbers.

\subsection{Action of $L_n (n>0)$ on Jack symmetric functions. Examples}

The relationship between the Fock space and the Jack symmetric
functions can be easily found as follows.
Let us first compare the dimensions of both systems. The degree-$n$ subspace of the
Fock space has dimension $p(n)$,
where $p(n)$ stand for the partition number, i.e. the number of ways to
express $n$ as a sum of all positive integers. On the other hand, number of
the degree-$n$ Jack symmetric function is equal to the number of Young
diagrams with $n$ boxes, i.e. $p(n)$. Then, from the orthogonality relation
(\ref{orthogonality}), the bosonized Jack symmetric functions form the
orthogonal basis of the Fock space,
\begin{equation}
  \mathcal{F}_A =\mathrm{span}\Bigl\{ J_{\lambda} |A\rangle \Bigl| \mathrm{any\,
  partition}\, \lambda \Bigl\} .
\end{equation}
In the same way, dual of the Fock space is generated by the Jack symmetric
functions as
\begin{equation}
  \mathcal{F}_A^* =\mathrm{span}\Bigl\{ \langle A|J_{\lambda} \Bigl| \mathrm{any\,
  partition}\, \lambda \Bigl\} .
\end{equation}

{}From now on, we consider the action of $L_n$ on the bosonized Jack
symmetric functions, i.e. the matrix representation of $L_n$ with respect
to the basis constructed by Jack symmetric functions. Before embarking on
the general formula, we first consider some examples and clarify the
combinatorial properties of the formula.

These formulas can be expressed as hooks on the Young diagrams. We prepare
some legends of diagrammatic explanations we will use later. We express the
hook within the Young diagram as broad line, and assign the term
$(m+n\beta)$ to the hook with horizontal length $m$ and vertical length
$n$. If there exist multiple hooks in one diagram, then we take product of
each term. 
We now give some examples.

\unitlength 10pt
\begin{picture}(30,4)(-5,-1)
\multiput(0,0)(1,0){2}{\line(0,1){2}}
\put(0,0){\line(1,0){1}}
\multiput(0,1)(0,1){2}{\line(1,0){3}}
\multiput(2,1)(1,0){2}{\line(0,1){1}}
\put(1,0){\circle*{0.5}}
\put(3,1){\circle*{0.5}}
\thicklines
\put(0.96,0){\line(0,1){1.07}}
\put(1.06,0){\line(0,1){1}}
\put(1,0.94){\line(1,0){2}}
\put(0.92,1.06){\line(1,0){2.05}}
\put(3.5,0.7){$=(2+\beta),$}
\thinlines
\multiput(9,0)(1,0){2}{\line(0,1){2}}
\put(9,0){\line(1,0){1}}
\multiput(9,1)(0,1){2}{\line(1,0){3}}
\multiput(11,1)(1,0){2}{\line(0,1){1}}
\put(12,1){\circle*{0.5}}
\put(12,2){\circle*{0.5}}
\thicklines
\put(11.94,1){\line(0,1){1}}
\put(12.06,1){\line(0,1){1}}
\put(12.5,0.7){$=(0+\beta),$}
\thinlines
\multiput(18,0)(1,0){2}{\line(0,1){2}}
\put(18,0){\line(1,0){1}}
\multiput(18,1)(0,1){2}{\line(1,0){3}}
\multiput(20,1)(1,0){2}{\line(0,1){1}}
\put(19,0){\circle*{0.5}}
\put(21,2){\circle*{0.5}}
\put(21,1){\circle*{0.5}}
\thicklines
\put(18.94,0){\line(0,1){2.08}}
\put(19.06,0){\line(0,1){2}}
\put(18.9,2.06){\line(1,0){2.1}}
\put(19,1.94){\line(1,0){2}}
\put(20.94,1){\line(0,1){1}}
\put(21.06,1){\line(0,1){1}}
\put(21.5,0.7){$=\beta\cdot (2+2\beta).$}
\end{picture}

We need another terminology to make our discussion simple enough.
A diagram being given, we define \textit{outer-corners} (black dots) and
\textit{inner-corners} (white dots) as in the following picture:

\begin{center}
\unitlength 10pt
\begin{picture}(10,5)
\put(0,0){\line(0,1){5}} \put(1,0){\line(0,1){1}} \put(2,1){\line(0,1){1}}
\put(3,2){\line(0,1){2}} \put(5,4){\line(0,1){1}}
\put(0,0){\line(1,0){1}} \put(1,1){\line(1,0){1}} \put(2,2){\line(1,0){1}}
\put(3,4){\line(1,0){2}} \put(0,5){\line(1,0){5}}
\put(1,0){\makebox(0,0){$\bullet$}} \put(2,1){\makebox(0,0){$\bullet$}}
\put(3,2){\makebox(0,0){$\bullet$}} \put(5,4){\makebox(0,0){$\bullet$}}
\put(0,0){\makebox(0,0){$\circ$}} \put(1,1){\makebox(0,0){$\circ$}}
\put(2,2){\makebox(0,0){$\circ$}} \put(3,4){\makebox(0,0){$\circ$}}
\put(5,5){\makebox(0,0){$\circ$}}
\end{picture}
\end{center}

First example is action of $L_1$ on $J
\unitlength 4pt
\begin{picture}(3.3,2)(0,1)
\multiput(0,0)(1,0){2}{\line(0,1){2}}
\put(0,0){\line(1,0){1}}
\multiput(0,1)(0,1){2}{\line(1,0){3}}
\multiput(2,1)(1,0){2}{\line(0,1){1}}
\end{picture}$.
\begin{eqnarray*}
\langle A_{r+1,s+1}|\, J
\unitlength 4pt
\begin{picture}(3.3,2)(0,1)
\multiput(0,0)(1,0){2}{\line(0,1){2}}
\put(0,0){\line(1,0){1}}
\multiput(0,1)(0,1){2}{\line(1,0){3}}
\multiput(2,1)(1,0){2}{\line(0,1){1}}
\end{picture}\,
L_1
&=&\langle A_{r+1,s+1}|\,\sqrt{2\beta}\\
& &\quad\,\,\times\left(
\frac{\beta\, (2+2\beta)}{(2+\beta)(3+2\beta)} \, A_{r-1,s-4} \, J
\unitlength 4pt
\begin{picture}(4,2)(0,1)
\multiput(0,0)(1,0){2}{\line(0,1){2}}
\put(0,0){\line(1,0){1}}
\multiput(0,1)(0,1){2}{\line(1,0){4}}
\multiput(2,1)(1,0){3}{\line(0,1){1}}
\end{picture}
\right.\\
\unitlength 10pt
\begin{picture}(10,5)(-18,0)
\multiput(0,0)(1,0){2}{\line(0,1){2}}
\put(0,0){\line(1,0){1}}
\multiput(0,1)(0,1){2}{\line(1,0){4}}
\multiput(2,1)(1,0){3}{\line(0,1){1}}
\put(3.25,1.15){1}
\put(-0.5,2.5){\line(1,0){5}}
\multiput(0,3)(1,0){2}{\line(0,1){2}}
\put(0,3){\line(1,0){1}}
\multiput(0,4)(0,1){2}{\line(1,0){4}}
\multiput(2,4)(1,0){3}{\line(0,1){1}}
\put(3.25,4.15){1}
\thicklines
\put(-0.04,0){\line(0,1){2.05}}
\put(0.04,0){\line(0,1){2.05}}
\put(0,1.96){\line(1,0){3}}
\put(-0.08,2.04){\line(1,0){3.08}}
\put(0.96,1){\line(0,1){1}}
\put(1.04,1){\line(0,1){1}}
\put(0,0){\circle*{0.5}}
\put(1,1){\circle*{0.5}}
\put(3,2){\circle*{0.5}}
\put(0.96,3){\line(0,1){2.08}}
\put(1.04,3){\line(0,1){2}}
\put(1,4.96){\line(1,0){2}}
\put(1,5.04){\line(1,0){2}}
\put(2.96,4){\line(0,1){1}}
\put(3.04,4){\line(0,1){1}}
\put(1,3){\circle*{0.5}}
\put(3,4){\circle*{0.5}}
\put(3,5){\circle*{0.5}}
\end{picture}\\
& &\qquad\,\,\,
+\frac{2\,\beta}{(2+\beta)(1+\beta)}\, A_{r-2,s-2}\,J
\unitlength 4pt
\begin{picture}(3,2)(0,1)
\multiput(0,0)(1,0){3}{\line(0,1){2}}
\put(0,0){\line(1,0){2}}
\multiput(0,1)(0,1){2}{\line(1,0){3}}
\put(3,1){\line(0,1){1}}
\end{picture}\\
\unitlength 10pt
\begin{picture}(10,5)(-18,0)
\multiput(0,0)(1,0){3}{\line(0,1){2}}
\put(0,0){\line(1,0){2}}
\multiput(0,1)(0,1){2}{\line(1,0){3}}
\put(3,1){\line(0,1){1}}
\put(1.25,0.15){1}
\put(-0.5,2,5){\line(1,0){4}}
\multiput(0,3)(1,0){3}{\line(0,1){2}}
\put(0,3){\line(1,0){2}}
\multiput(0,4)(0,1){2}{\line(1,0){3}}
\put(3,4){\line(0,1){1}}
\put(1.25,3.15){1}
\thicklines
\put(-0.04,0){\line(0,1){1.07}}
\put(0.04,0){\line(0,1){1}}
\put(0,0.96){\line(1,0){1}}
\put(0,1.04){\line(1,0){1}}
\put(0.96,1){\line(0,1){1.08}}
\put(1.04,1){\line(0,1){1}}
\put(1,1.96){\line(1,0){2}}
\put(1,2.04){\line(1,0){2}}
\put(0,0){\circle*{0.5}}
\put(1,1){\circle*{0.5}}
\put(3,2){\circle*{0.5}}
\put(0.96,3){\line(0,1){1}}
\put(1.04,3){\line(0,1){1}}
\put(1,3.96){\line(1,0){2}}
\put(1,4.04){\line(1,0){2}}
\put(1,3){\circle*{0.5}}
\put(1,4){\circle*{0.5}}
\put(3,4){\circle*{0.5}}
\end{picture}\\
& &\qquad\,\,\,\left.
+\frac{1\, (3+\beta)}{(1+\beta)(3+2\beta)}\,
A_{r-3,s-1}\, J
\unitlength 4pt
\begin{picture}(3,3)(0,2)
\multiput(0,0)(1,0){2}{\line(0,1){3}}
\multiput(0,0)(0,1){2}{\line(1,0){1}}
\multiput(0,2)(0,1){2}{\line(1,0){3}}
\multiput(2,2)(1,0){2}{\line(0,1){1}}
\end{picture}\right) \\
\unitlength 10pt
\begin{picture}(10,7)(-18,0)
\multiput(0,0)(1,0){2}{\line(0,1){3}}
\multiput(0,0)(0,1){2}{\line(1,0){1}}
\multiput(0,2)(0,1){2}{\line(1,0){3}}
\multiput(2,2)(1,0){2}{\line(0,1){1}}
\put(0.25,0.15){1}
\put(-0.5,3.5){\line(1,0){4}}
\multiput(0,4)(1,0){2}{\line(0,1){3}}
\multiput(0,4)(0,1){2}{\line(1,0){1}}
\multiput(0,6)(0,1){2}{\line(1,0){3}}
\multiput(2,6)(1,0){2}{\line(0,1){1}}
\put(0.25,4.15){1}
\thicklines
\put(-0.04,1){\line(0,1){2.08}}
\put(0.04,1){\line(0,1){2}}
\put(0,1.96){\line(1,0){1}}
\put(0,2.04){\line(1,0){1}}
\put(0,2.96){\line(1,0){3}}
\put(0,3.04){\line(1,0){3}}
\put(0,1){\circle*{0.5}}
\put(1,2){\circle*{0.5}}
\put(3,3){\circle*{0.5}}
\put(0,4.96){\line(1,0){1}}
\put(0,5.04){\line(1,0){1}}
\put(-0.04,5){\line(0,1){1.08}}
\put(0.04,5){\line(0,1){1}}
\put(0,5.96){\line(1,0){3}}
\put(0,6.04){\line(1,0){3}}
\put(0,5){\circle*{0.5}}
\put(1,5){\circle*{0.5}}
\put(3,6){\circle*{0.5}}
\end{picture}
\end{eqnarray*}
In above equation, diagrams are inserted for explanation of rational
functions of $\beta$. Two diagrams beside the horizontal line stand for
numerators and denominators respectively. In the numerator, we join upper
left corner of box ``1" and all the outer-corners of \unitlength 4pt
\thinlines \begin{picture}(3.3,2)(0,0.5)
\multiput(0,0)(1,0){2}{\line(0,1){2}} \put(0,0){\line(1,0){1}}
\multiput(0,1)(0,1){2}{\line(1,0){3}} \multiput(2,1)(1,0){2}{\line(0,1){1}}
\end{picture}
by hook. On the other hand, in the denominator, we join upper left corner
of box ``1" and all the inner-corners of 
\unitlength 4pt 
\thinlines
\begin{picture}(3.3,2)(0,0.5) \multiput(0,0)(1,0){2}{\line(0,1){2}}
\put(0,0){\line(1,0){1}} \multiput(0,1)(0,1){2}{\line(1,0){3}}
\multiput(2,1)(1,0){2}{\line(0,1){1}} \end{picture}. If added box -- box
``1"-- has coordinate $(i,j)$, then the term $A_{r-i,s-j}$ appears.

As a second example, we show the action of $L_2$.
\pagebreak
\begin{eqnarray*}
\langle A_{r+1,s+1}|J
\unitlength 4pt
\begin{picture}(3.3,2)(0,1)
\multiput(0,0)(1,0){2}{\line(0,1){2}}
\put(0,0){\line(1,0){1}}
\multiput(0,1)(0,1){2}{\line(1,0){3}}
\multiput(2,1)(1,0){2}{\line(0,1){1}}
\end{picture}\, L_2
&=&\langle A_{r+1,s+1} |\sqrt{2\beta}\,\beta\\
& &
\times\left( \frac{\beta\, (2+2\beta)}{(2+\beta)(3+2\beta)}
\times\frac{(3+2\beta)}{(3+\beta)(4+2\beta)}
A_{r-1,s-5}J
\unitlength 4pt
\begin{picture}(5,2)(0,1)
\multiput(0,0)(1,0){2}{\line(0,1){2}}
\put(0,0){\line(1,0){1}}
\multiput(0,1)(0,1){2}{\line(1,0){5}}
\multiput(2,1)(1,0){4}{\line(0,1){1}}
\end{picture}
\right.\\
\unitlength 10pt
\begin{picture}(10,5)(-18,0)
\multiput(0,0)(1,0){2}{\line(0,1){2}}
\put(0,0){\line(1,0){1}}
\multiput(0,1)(0,1){2}{\line(1,0){4}}
\multiput(2,1)(1,0){3}{\line(0,1){1}}
\put(-0.5,2.5){\line(1,0){5}}
\multiput(0,3)(1,0){2}{\line(0,1){2}}
\put(0,3){\line(1,0){1}}
\multiput(0,4)(0,1){2}{\line(1,0){4}}
\multiput(2,4)(1,0){3}{\line(0,1){1}}
\put(3.25,1.15){1}
\put(3.25,4.15){1}
\thicklines
\put(-0.04,0){\line(0,1){2.08}}
\put(0.04,0){\line(0,1){2}}
\put(0,1.96){\line(1,0){3}}
\put(0,2.04){\line(1,0){3}}
\put(0.96,1){\line(0,1){1}}
\put(1.04,1){\line(0,1){1}}
\put(0,0){\circle*{0.5}}
\put(1,1){\circle*{0.5}}
\put(3,2){\circle*{0.5}}
\put(0.96,3){\line(0,1){2.08}}
\put(1.04,3){\line(0,1){2}}
\put(1,4.96){\line(1,0){2}}
\put(1,5.04){\line(1,0){2}}
\put(2.96,4){\line(0,1){1}}
\put(3.04,4){\line(0,1){1}}
\put(1,3){\circle*{0.5}}
\put(3,4){\circle*{0.5}}
\put(3,5){\circle*{0.5}}
\put(5.7,2.15){$\times$}
\thinlines
\multiput(8,0)(1,0){2}{\line(0,1){2}}
\put(8,0){\line(1,0){1}}
\multiput(8,1)(0,1){2}{\line(1,0){5}}
\multiput(10,1)(1,0){4}{\line(0,1){1}}
\put(7.5,2.5){\line(1,0){6}}
\multiput(8,3)(1,0){2}{\line(0,1){2}}
\put(8,3){\line(1,0){1}}
\multiput(8,4)(0,1){2}{\line(1,0){5}}
\multiput(10,4)(1,0){4}{\line(0,1){1}}
\put(11.25,1.15){1}
\put(12.25,1.15){2}
\put(11.25,4.15){1}
\put(12.25,4.15){2}
\thicklines
\put(7.96,0){\line(0,1){2.08}}
\put(8.04,0){\line(0,1){2}}
\put(8,1.96){\line(1,0){4}}
\put(8,2.04){\line(1,0){4}}
\put(8.96,1){\line(0,1){1}}
\put(9.04,1){\line(0,1){1}}
\put(8,0){\circle*{0.5}}
\put(9,1){\circle*{0.5}}
\put(12,2){\circle*{0.5}}
\put(8.96,3){\line(0,1){2.08}}
\put(9.04,3){\line(0,1){2}}
\put(9,4.96){\line(1,0){3}}
\put(9,5.04){\line(1,0){3}}
\put(9,3){\circle*{0.5}}
\put(12,5){\circle*{0.5}}
\end{picture}
\\
& &\quad
-\frac{\beta\, (2+2\beta)}{(2+\beta)(3+2\beta)}\times
\frac{\beta}{(3+\beta)(1+\beta)}
A_{r-2,s-2}\, J
\unitlength 4pt
\begin{picture}(4,2)(0,1)
\multiput(0,0)(1,0){3}{\line(0,1){2}}
\put(0,0){\line(1,0){2}}
\multiput(0,1)(0,1){2}{\line(1,0){4}}
\multiput(3,1)(1,0){2}{\line(0,1){1}}
\end{picture}\\
\unitlength 10pt
\begin{picture}(10,5)(-18,0)
\multiput(0,0)(1,0){2}{\line(0,1){2}}
\put(0,0){\line(1,0){1}}
\multiput(0,1)(0,1){2}{\line(1,0){4}}
\multiput(2,1)(1,0){3}{\line(0,1){1}}
\put(3.25,1.15){1}
\put(-0.5,2.5){\line(1,0){5}}
\multiput(0,3)(1,0){2}{\line(0,1){2}}
\put(0,3){\line(1,0){1}}
\multiput(0,4)(0,1){2}{\line(1,0){4}}
\multiput(2,4)(1,0){3}{\line(0,1){1}}
\put(3.25,4.15){1}
\thicklines
\put(-0.04,0){\line(0,1){2.05}}
\put(0.04,0){\line(0,1){2.05}}
\put(0,1.96){\line(1,0){3}}
\put(-0.08,2.04){\line(1,0){3.08}}
\put(0.96,1){\line(0,1){1}}
\put(1.04,1){\line(0,1){1}}
\put(0,0){\circle*{0.5}}
\put(1,1){\circle*{0.5}}
\put(3,2){\circle*{0.5}}
\put(0.96,3){\line(0,1){2.08}}
\put(1.04,3){\line(0,1){2}}
\put(1,4.96){\line(1,0){2}}
\put(1,5.04){\line(1,0){2}}
\put(2.96,4){\line(0,1){1}}
\put(3.04,4){\line(0,1){1}}
\put(1,3){\circle*{0.5}}
\put(3,4){\circle*{0.5}}
\put(3,5){\circle*{0.5}}
\put(5.7,2.15){$\times$}
\thinlines
\multiput(8,0)(1,0){3}{\line(0,1){2}}
\put(8,0){\line(1,0){2}}
\multiput(8,1)(0,1){2}{\line(1,0){4}}
\multiput(10,1)(1,0){3}{\line(0,1){1}}
\put(11.25,1.15){1}
\put(9.25,0.15){2}
\put(7.5,2.5){\line(1,0){5}}
\multiput(8,3)(1,0){3}{\line(0,1){2}}
\put(8,3){\line(1,0){2}}
\multiput(8,4)(0,1){2}{\line(1,0){4}}
\multiput(10,4)(1,0){3}{\line(0,1){1}}
\put(11.25,4.15){1}
\put(9.25,3.15){2}
\thicklines
\put(7.96,0){\line(0,1){1.05}}
\put(8.04,0){\line(0,1){1.05}}
\put(9,1.96){\line(1,0){3}}
\put(8.92,2.04){\line(1,0){3.08}}
\put(8.96,1){\line(0,1){1}}
\put(9.04,1){\line(0,1){1}}
\put(8,0.96){\line(1,0){1}}
\put(7.92,1.04){\line(1,0){1.04}}
\put(8,0){\circle*{0.5}}
\put(9,1){\circle*{0.5}}
\put(12,2){\circle*{0.5}}
\put(8.96,3){\line(0,1){1.04}}
\put(9.04,3){\line(0,1){1}}
\put(9,3){\circle*{0.5}}
\put(9,4){\circle*{0.5}}
\end{picture}
\\
& &\quad
+\frac{2\beta}{(2+\beta)(1+\beta)}\times
\frac{\beta}{(1+\beta)(3+2\beta)}\,
A_{r-1,s-4}\, J
\unitlength 4pt
\begin{picture}(4,2)(0,1)
\multiput(0,0)(1,0){3}{\line(0,1){2}}
\put(0,0){\line(1,0){2}}
\multiput(0,1)(0,1){2}{\line(1,0){4}}
\multiput(3,1)(1,0){2}{\line(0,1){1}}
\end{picture}\\
\unitlength 10pt
\begin{picture}(10,5)(-18,0)
\multiput(0,0)(1,0){3}{\line(0,1){2}}
\put(0,0){\line(1,0){2}}
\multiput(0,1)(0,1){2}{\line(1,0){3}}
\multiput(2,1)(1,0){2}{\line(0,1){1}}
\put(1.25,0.15){1}
\put(-0.5,2.5){\line(1,0){5}}
\multiput(0,3)(1,0){3}{\line(0,1){2}}
\put(0,3){\line(1,0){2}}
\multiput(0,4)(0,1){2}{\line(1,0){3}}
\multiput(2,4)(1,0){2}{\line(0,1){1}}
\put(1.25,3.15){1}
\thicklines
\put(-0.04,0){\line(0,1){1.04}}
\put(0.04,0){\line(0,1){1}}
\put(0,0.96){\line(1,0){1}}
\put(-0.08,1.04){\line(1,0){1.04}}
\put(1,1.96){\line(1,0){2}}
\put(0.92,2.04){\line(1,0){2.08}}
\put(0.96,1){\line(0,1){1}}
\put(1.04,1){\line(0,1){1}}
\put(0,0){\circle*{0.5}}
\put(1,1){\circle*{0.5}}
\put(3,2){\circle*{0.5}}
\put(0.96,3){\line(0,1){1.04}}
\put(1.04,3){\line(0,1){1}}
\put(1,3.96){\line(1,0){2}}
\put(1,4.04){\line(1,0){2}}
\put(1,3){\circle*{0.5}}
\put(1,4){\circle*{0.5}}
\put(3,4){\circle*{0.5}}
\put(5.7,2.15){$\times$}
\thinlines
\multiput(8,0)(1,0){3}{\line(0,1){2}}
\put(8,0){\line(1,0){2}}
\multiput(8,1)(0,1){2}{\line(1,0){4}}
\multiput(10,1)(1,0){3}{\line(0,1){1}}
\put(11.25,1.15){2}
\put(9.25,0.15){1}
\put(7.5,2.5){\line(1,0){5}}
\multiput(8,3)(1,0){3}{\line(0,1){2}}
\put(8,3){\line(1,0){2}}
\multiput(8,4)(0,1){2}{\line(1,0){4}}
\multiput(10,4)(1,0){3}{\line(0,1){1}}
\put(11.25,4.15){2}
\put(9.25,3.15){1}
\thicklines
\put(7.96,0){\line(0,1){2.05}}
\put(8.04,0){\line(0,1){2.05}}
\put(8,1.96){\line(1,0){3}}
\put(7.92,2.04){\line(1,0){3.08}}
\put(9.96,1){\line(0,1){1}}
\put(10.04,1){\line(0,1){1}}
\put(8,0){\circle*{0.5}}
\put(10,1){\circle*{0.5}}
\put(11,2){\circle*{0.5}}
\put(10.96,4){\line(0,1){1}}
\put(11.04,4){\line(0,1){1}}
\put(11,4){\circle*{0.5}}
\put(11,5){\circle*{0.5}}
\end{picture}
\\
& &\quad
-\frac{\beta(2+2\beta)}{(2+\beta)(3+2\beta)}\times
\frac{1}{(1+\beta)(4+2\beta)}\,
A_{r-3,s-1}\, J
\unitlength 4pt
\begin{picture}(4,2)(0,2)
\multiput(0,0)(1,0){2}{\line(0,1){3}}
\multiput(0,0)(0,1){2}{\line(1,0){1}}
\multiput(0,2)(0,1){2}{\line(1,0){4}}
\multiput(2,2)(1,0){3}{\line(0,1){1}}
\end{picture}\\
& &\\
& &\quad
+\frac{3+\beta}{(1+\beta)(3+2\beta)}\times
\frac{\beta}{(2+\beta)(3+3\beta)}\,
A_{r-1,s-4}\, J
\unitlength 4pt
\begin{picture}(4,2)(0,2)
\multiput(0,0)(1,0){2}{\line(0,1){3}}
\multiput(0,0)(0,1){2}{\line(1,0){1}}
\multiput(0,2)(0,1){2}{\line(1,0){4}}
\multiput(2,2)(1,0){3}{\line(0,1){1}}
\end{picture}\\
& &\\
& &\quad
-\frac{2\beta}{(1+\beta)(2+\beta)}\times
\frac{3+\beta}{(2+\beta)(3+2\beta)}\,
A_{r-3,s-1}\, J
\unitlength 4pt
\begin{picture}(4,2)(0,2)
\multiput(0,0)(1,0){2}{\line(0,1){3}}
\put(0,0){\line(1,0){1}}
\put(0,1){\line(1,0){2}}
\multiput(0,2)(0,1){2}{\line(1,0){3}}
\put(2,1){\line(0,1){2}}
\put(3,2){\line(0,1){1}}
\end{picture}\\
& &\\
& &\quad
+\frac{3+\beta}{(1+\beta)(3+2\beta)}\times
\frac{2}{(2+\beta)(1+2\beta)}\,
A_{r-2,s-2}\, J
\unitlength 4pt
\begin{picture}(4,2)(0,2)
\multiput(0,0)(1,0){2}{\line(0,1){3}}
\put(0,0){\line(1,0){1}}
\put(0,1){\line(1,0){2}}
\multiput(0,2)(0,1){2}{\line(1,0){3}}
\put(2,1){\line(0,1){2}}
\put(3,2){\line(0,1){1}}
\end{picture}\\
& &\\
& &\quad
+\frac{2\beta}{(1+\beta)(2+\beta)}\times
\frac{1}{(1+\beta)(2+\beta)}\,
A_{r-2,s-3}\, J
\unitlength 4pt
\begin{picture}(3,2)(0,1)
\multiput(0,0)(1,0){4}{\line(0,1){2}}
\multiput(0,0)(0,1){3}{\line(1,0){3}}
\end{picture}\\
& &\\
& &\quad
-\frac{3+\beta}{(1+\beta)(3+2\beta)}\times
\frac{3+2\beta}{(3+3\beta)(1+2\beta)}\,
A_{r-4,s-1}\, J
\unitlength 4pt
\begin{picture}(3,3)(0,3)
\multiput(0,0)(1,0){2}{\line(0,1){4}}
\multiput(0,0)(0,1){3}{\line(1,0){1}}
\multiput(0,3)(0,1){2}{\line(1,0){3}}
\multiput(2,3)(1,0){2}{\line(0,1){1}}
\end{picture}
\;\;\bigg)
\end{eqnarray*}
In this case, graphical meaning of the rational function of $\beta$ is
almost the same as in the action of $L_1$. Note, however, we don't take the corner
created by the previously added box in the numerator. If the last added box --box
``2"-- has coordinate $(i,j)$, then factor $A_{r-i,s-j}$ appear. When row
of box ``1" is lesser than that of box ``2", then a minus sign appears.

\subsection{Action of $L_n$: general formula}

We take a general Young diagram $Y$ and $n>0$, and consider $\langle
A_{r+1,s+1} |J_Y L_n$. We parameterize $Y$ as
\begin{equation}
Y=Y^{(1)}=\left( {s_1^{(1)}}^{r_1^{(1)} -r_0^{(1)}},
{s_2^{(1)}}^{r_2^{(1)} -r_1^{(1)}},\cdots ,
{s_{m^{(1)}}^{(1)}}^{r_{m^{(1)}}^{(1)}-r_{m^{(1)}-1}^{(1)}}
\right) . \label{para}
\end{equation}
We take $r_0^{(1)}=0$, $s_{m^{(1)}+1}^{(1)}=0$. In this parameterization,
the outer-corners of $Y^{(1)}$ are given by $(r_1^{(1)},s_1^{(1)})$,
$(r_2^{(1)},s_2^{(1)})$, $\cdots$, $(r_{m^{(1)}}^{(1)},s_{m^{(1)}}^{(1)})$,
where symbol ``$(1)$" indicate that this diagram is going to be
added by the first box.

\textbf{General formula}

\textbf{Step 1:} 
Add one box, say to the place $(r_{i_1 -1}^{(1)}+1,s_{i_1}^{(1)}+1)$,
 to $Y^{(1)}$, and denote it as
$Y^{(1)}\cup(r_{i_1 -1}^{(1)}+1,s_{i_1}^{(1)}+1)$. Associate a coefficient to $Y^{(1)}$,
which is given by
\begin{eqnarray}
  & & \prod _{j=1}^{i_1 -1} \frac{\left[\left(
  a(r_j^{(1)},s_{i_1}^{(1)}+1)+1\right)
  +l(r_j^{(1)},s_{i_1}^{(1)}+1)\beta\, \right]}{ \left[\left(
  a(r_{j-1}^{(1)}+1,s_{i_1}^{(1)}+1)+1\right) +\left(
  l(r_{j-1}^{(1)}+1,s_{i_1}^{(1)} +1)+1\right) \beta\, \right] } \\
  & & \times\prod_{j=i_1}^{m^{(1)}} \frac{\left[ a(r_{i_1
  -1}^{(1)}+1,s_j^{(1)})+ \left( l(r_{i_1 -1}^{(1)}+1,s_j^{(1)})+1\right)
  \beta\,\right] }{ \left[\left( a(r_{i_1
  -1}^{(1)}+1,s_{j+1}^{(1)}+1)+1\right) +\left( l(r_{i_1
  -1}^{(1)}+1,s_{j+1}^{(1)}+1)+1\right) \beta\,\right] }.\nonumber
\end{eqnarray}
To make the meaning of the coordinates used in the above formula clearer, 
we give a schematic interpretation below.
In this diagram, upper left diagram correspond to the numerator of the first term in above 
formula, and so on.

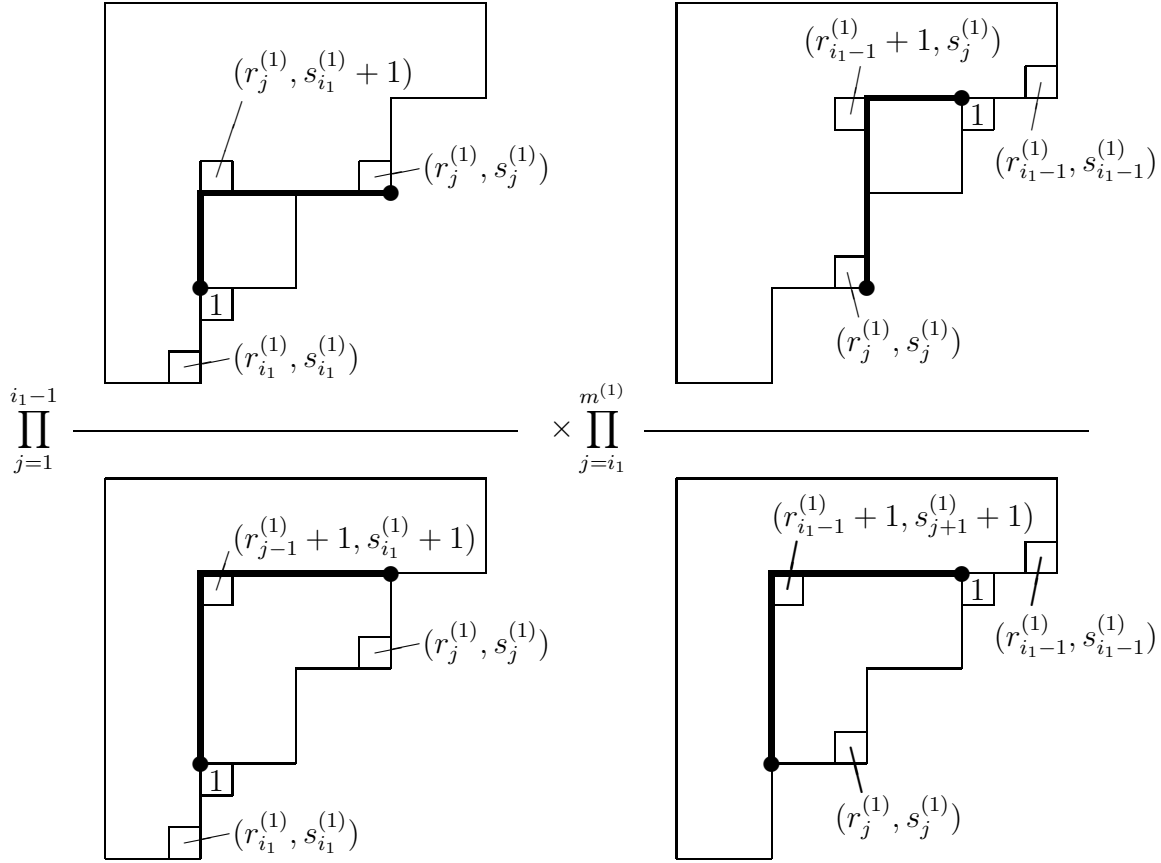
\unitlength 12pt
\begin{picture}(30,28)(-2,0)
\put(0,0){\line(1,0){3}}
\put(0,0){\line(0,1){12}}
\put(3,0){\line(0,1){3}}
\put(3,3){\line(1,0){3}}
\put(6,3){\line(0,1){3}}
\put(6,6){\line(1,0){3}}
\put(9,6){\line(0,1){3}}
\put(9,9){\line(1,0){3}}
\put(12,9){\line(0,1){3}}
\put(12,12){\line(-1,0){12}}
\put(3,3){\circle*{0.5}}
\put(9,9){\circle*{0.5}}
\put(2,0){\line(0,1){1}}
\put(2,1){\line(1,0){1}}
\put(8,6){\line(0,1){1}}
\put(8,7){\line(1,0){1}}
\put(3,8){\line(1,0){1}}
\put(4,8){\line(0,1){1}}
\put(3,2){\line(1,0){1}}
\put(4,2){\line(0,1){1}}
\put(3.25,2.15){1}
\thicklines
\put(2.95,3){\line(0,1){6.09}}
\put(3,3){\line(0,1){6.09}}
\put(3.05,3){\line(0,1){6}}
\put(3,8.95){\line(1,0){6}}
\put(3,9){\line(1,0){6}}
\put(3,9.05){\line(1,0){6}}
\thinlines
\put(2.5,0.5){\line(5,1){1.4}}
\put(4,0.5){$(r_{i_1}^{(1)},s_{i_1}^{(1)})$}
\put(8.5,6.5){\line(5,1){1.4}}
\put(10,6.5){$(r_j^{(1)},s_j^{(1)})$}
\put(3.5,8.5){\line(1,3){0.4}}
\put(4,9.8){$(r_{j-1}^{(1)}+1,s_{i_1}^{(1)}+1)$}
\put(-1,13.5){\line(1,0){14}}
\put(-3,13.3){$\displaystyle \prod_{j=1}^{i_1 -1}$}

\put(0,15){\line(1,0){3}}
\put(0,15){\line(0,1){12}}
\put(3,15){\line(0,1){3}}
\put(3,18){\line(1,0){3}}
\put(6,18){\line(0,1){3}}
\put(6,21){\line(1,0){3}}
\put(9,21){\line(0,1){3}}
\put(9,24){\line(1,0){3}}
\put(12,24){\line(0,1){3}}
\put(12,27){\line(-1,0){12}}
\put(2,15){\line(0,1){1}}
\put(2,16){\line(1,0){1}}
\put(8,21){\line(0,1){1}}
\put(8,22){\line(1,0){1}}
\put(3,21){\line(0,1){1}}
\put(3,22){\line(1,0){1}}
\put(4,22){\line(0,-1){1}}
\put(3,17){\line(1,0){1}}
\put(4,17){\line(0,1){1}}
\put(3.25,17.15){1}
\thicklines
\put(2.95,18){\line(0,1){3.08}}
\put(3,18){\line(0,1){3}}
\put(3.05,18){\line(0,1){3}}
\put(3,20.95){\line(1,0){6}}
\put(3,21){\line(1,0){6}}
\put(3,21.05){\line(1,0){6}}
\put(3,18){\circle*{0.5}}
\put(9,21){\circle*{0.5}}
\thinlines
\put(2.5,15.5){\line(5,1){1.4}}
\put(4,15.5){$(r_{i_1}^{(1)},s_{i_1}^{(1)})$}
\put(3.5,21.5){\line(1,3){0.8}}
\put(4,24.5){$(r_j^{(1)},s_{i_1}^{(1)}+1)$}
\put(8.5,21.5){\line(5,1){1.4}}
\put(10,21.5){$(r_j^{(1)},s_j^{(1)})$}

\put(14,13.3){$\times \displaystyle\prod_{j=i_1}^{m^{(1)}}$}

\put(18,0){\line(0,1){12}}
\put(18,0){\line(1,0){3}}
\put(21,0){\line(0,1){3}}
\put(21,3){\line(1,0){3}}
\put(24,3){\line(0,1){3}}
\put(24,6){\line(1,0){3}}
\put(27,6){\line(0,1){3}}
\put(27,9){\line(1,0){3}}
\put(30,9){\line(0,1){3}}
\put(30,12){\line(-1,0){12}}
\put(17,13.5){\line(1,0){14}}
\put(23,3){\line(0,1){1}}
\put(23,4){\line(1,0){1}}
\put(27,8){\line(1,0){1}}
\put(28,8){\line(0,1){1}}
\put(27.25,8.15){1}
\put(29,9){\line(0,1){1}}
\put(29,10){\line(1,0){1}}
\put(21,8){\line(1,0){1}}
\put(22,8){\line(0,1){1}}
\put(21,3){\circle*{0.5}}
\put(27,9){\circle*{0.5}}
\thicklines
\put(20.95,3){\line(0,1){6.09}}
\put(21,3){\line(0,1){6.09}}
\put(21.05,3){\line(0,1){6}}
\put(21,8.95){\line(1,0){6}}
\put(21,9){\line(1,0){6}}
\put(21,9.05){\line(1,0){6}}
\put(23.5,3.5){\line(1,-4){0.35}}
\put(23,1){$(r_j^{(1)},s_j^{(1)})$}
\put(21,10.5){$(r_{i_1 -1}^{(1)}+1,s_{j+1}^{(1)}+1)$}
\put(21.5,8.5){\line(1,5){0.3}}
\put(29.5,9.5){\line(-1,-5){0.3}}
\put(28,6.7){$(r_{i_1-1}^{(1)},s_{i_1-1}^{(1)})$}

\thinlines
\put(18,15){\line(0,1){12}}
\put(18,15){\line(1,0){3}}
\put(21,15){\line(0,1){3}}
\put(21,18){\line(1,0){3}}
\put(24,18){\line(0,1){3}}
\put(24,21){\line(1,0){3}}
\put(27,21){\line(0,1){3}}
\put(27,24){\line(1,0){3}}
\put(30,24){\line(0,1){3}}
\put(30,27){\line(-1,0){12}}
\put(23,18){\line(0,1){1}}
\put(23,19){\line(1,0){1}}
\put(23,23){\line(1,0){1}}
\put(23,23){\line(0,1){1}}
\put(23,24){\line(1,0){1}}
\put(29,24){\line(0,1){1}}
\put(29,25){\line(1,0){1}}
\put(24,18){\circle*{0.5}}
\put(27,24){\circle*{0.5}}
\put(27,23){\line(1,0){1}}
\put(28,23){\line(0,1){1}}
\put(27.25,23.15){1}
\thicklines
\put(23.95,18){\line(0,1){6.08}}
\put(24,18){\line(0,1){6.08}}
\put(24.05,18){\line(0,1){6}}
\put(24,23.95){\line(1,0){3}}
\put(24,24){\line(1,0){3}}
\put(24,24.05){\line(1,0){3}}
\thinlines
\put(23.4,18.5){\line(1,-4){0.35}}
\put(23,16){$(r_j^{(1)},s_j^{(1)})$}
\put(22,25.5){$(r_{i_1-1}^{(1)}+1,s_j^{(1)})$}
\put(23.5,23.5){\line(1,5){0.3}}
\put(29.5,24.5){\line(-1,-5){0.3}}
\put(28,21.7){$(r_{i_1-1}^{(1)},s_{i_1-1}^{(1)})$}

\end{picture}
\bigskip

\textbf{Step 2:}
Take a coordinate of $Y^{(2)}=Y^{(1)}\cup (r_{i_1 -1}^{(1)}+1,s_{i_1 }^{(1)}+1)$ 
as in $Y^{(1)}$, namely, set
\begin{equation}
  Y^{(2)}=
  Y^{(1)}\cup (r_{i_1 -1}^{(1)}+1,s_{i_1 }^{(1)}+1) = \left( {s_1^{(2)}}^{r_1^{(2)}
  -r_0^{(2)}}, {s_2^{(2)}}^{r_2^{(2)} -r_1^{(2)}},\cdots ,
  {s_{m^{(2)}}^{(2)}}^{r_{m^{(2)}}^{(2)}-r_{m^{(2)}-1}^{(2)}} \right) .
\end{equation}
Add one more box to anywhere you want to this diagram if this addition gives us a Young diagram, 
and we obtain a similar factor as in Step 1.
In this step, however, we need to work with the following ``exception rule".
If the outer-corner $(r_k^{(2)},s_k^{(2)})$ is produced by the box just added in
the last step, we should omit the factor corresponding to
$(r_k^{(2)},s_k^{(2)})$ in the numerator.

\textbf{Step 3:}
Denote the Young diagram after the second addition as $Y^{(3)}
=Y^{(2)}\cup (r_{i_2 -1}^{(2)}+1,s_{i_2}^{(2)}+1)$.
Add third box to $Y^{(3)}$ and obtain similar factors as in Step 2.
Note that we need to work with the exception rule.
Repeat this manipulation recursively until $n$-th box is added.

\textbf{Step 4:} 
Multiply  
\begin{equation}
\beta^{n-1} (-1)^{\#\{ k|r_{i_k -1}^{(k)}<r_{i_{k+1}-1}^{(k+1)}\}}
J_{Y\cup (r_{i_1 -1}^{(1)} +1,s_{i_1}^{(1)}+1)\cup\cdots\cup
(r_{i_n -1}^{(n)}+1,s_{i_n}^{(n)}+1)},
\end{equation}
to the result of Step 3.

\textbf{Step 5:}
Multiply  
\begin{equation}
\langle A_{r+1,s+1}|
\sqrt{2\beta}A_{r-(r_{i_n -1}^{(n)}+1),s-(s_{i_n -1}^{(n)}+1)},
\end{equation}
to the result of Step 4.

\textbf{Step 6:} 
Repeat Steps 1 to 5 for each way to add $n$ boxes to $Y$, and sum up all
the terms.

\begin{remark}
  Formula for $p_n J_Y$ is obtained by doing steps 1 to 4 and 6.
\end{remark}

\subsection{Action of $L_{-n} (n>0)$ on Jack symmetric functions}

Before giving a general formula, we show an example.
\begin{eqnarray*}
\langle A_{r+1,s+1}|J
\unitlength 4pt
\begin{picture}(3,2)(0,1)
\multiput(0,0)(1,0){2}{\line(0,1){2}}
\put(0,0){\line(1,0){1}}
\multiput(0,1)(0,1){2}{\line(1,0){3}}
\multiput(2,1)(1,0){2}{\line(0,1){1}}
\end{picture}\, L_{-1} &=&
\langle A_{r+1,s+1}|
\frac{1}{\sqrt{2\beta}}\cdot \frac{1}{\beta}\\
& &\times\left(
\frac{(2+2\beta)\beta \cdot 1}{(2+\beta)}
A_{r+3-(2+2),s+3-(1+1)}J
\unitlength 4pt
\begin{picture}(3,2)(0,1)
\multiput(0,0)(0,1){2}{\line(1,0){3}}
\multiput(0,0)(1,0){4}{\line(0,1){1}}
\end{picture}\right.\\
\unitlength 10pt
\begin{picture}(3,5)(-8,0)
\multiput(0,0)(1,0){2}{\line(0,1){2}}
\put(0,0){\line(1,0){1}}
\multiput(0,1)(0,1){2}{\line(1,0){3}}
\multiput(2,1)(1,0){2}{\line(0,1){1}}
\put(-0.5,2.5){\line(1,0){4}}
\multiput(0,3)(1,0){2}{\line(0,1){2}}
\put(0,3){\line(1,0){1}}
\multiput(0,4)(0,1){2}{\line(1,0){3}}
\multiput(2,4)(1,0){2}{\line(0,1){1}}
\put(0.25,0.15){1}
\put(0.25,3.15){1}
\thicklines
\put(0.96,0){\line(0,1){1.08}}
\put(1.04,0){\line(0,1){1}}
\put(1,0.96){\line(1,0){2}}
\put(1,1.04){\line(1,0){2}}
\put(1,0){\circle*{0.5}}
\put(3,1){\circle*{0.5}}
\put(0,2.96){\line(1,0){1}}
\put(0,3.04){\line(1,0){1}}
\put(0.96,3){\line(0,1){2.08}}
\put(1.04,3){\line(0,1){2}}
\put(1,4.96){\line(1,0){2}}
\put(1,5.04){\line(1,0){2}}
\put(0,3){\circle*{0.5}}
\put(1,3){\circle*{0.5}}
\put(1,4){\circle*{0.5}}
\put(3,5){\circle*{0.5}}
\end{picture}\\
& &\left. +\frac{\beta\cdot 2\, (3+\beta)}{(2+\beta)}
A_{r+3-(1+1),s+3-(3+3)}J
\unitlength 4pt
\begin{picture}(2,2)(0,1)
\multiput(0,0)(1,0){2}{\line(0,1){2}}
\put(0,0){\line(1,0){1}}
\multiput(0,1)(0,1){2}{\line(1,0){2}}
\put(2,1){\line(0,1){1}}
\end{picture}\right)\\
\unitlength 10pt
\begin{picture}(3,5)(-8,0)
\multiput(0,0)(1,0){2}{\line(0,1){2}}
\put(0,0){\line(1,0){1}}
\multiput(0,1)(0,1){2}{\line(1,0){3}}
\multiput(2,1)(1,0){2}{\line(0,1){1}}
\put(-0.5,2.5){\line(1,0){4}}
\multiput(0,3)(1,0){2}{\line(0,1){2}}
\put(0,3){\line(1,0){1}}
\multiput(0,4)(0,1){2}{\line(1,0){3}}
\multiput(2,4)(1,0){2}{\line(0,1){1}}
\put(2.25,1.15){1}
\put(2.25,4.15){1}
\thicklines
\put(0.96,0){\line(0,1){1.08}}
\put(1.04,0){\line(0,1){1}}
\put(1,0.96){\line(1,0){2}}
\put(1,1.04){\line(1,0){2}}
\put(1,0){\circle*{0.5}}
\put(3,1){\circle*{0.5}}
\put(-0.04,3){\line(0,1){1.08}}
\put(0.04,3){\line(0,1){1}}
\put(0,3.96){\line(1,0){3}}
\put(0,4.04){\line(1,0){3}}
\put(2.96,4){\line(0,1){1}}
\put(3.04,4){\line(0,1){1}}
\put(0,3){\circle*{0.5}}
\put(1,4){\circle*{0.5}}
\put(3,4){\circle*{0.5}}
\put(3,5){\circle*{0.5}}
\end{picture}
\end{eqnarray*}
In the above diagram, the letter ``1" stands for the first removed box.
In the numerator, in contrast to the action of $L_1$, we join lower right 
corner of box ``1" with all inner corners of 
\unitlength 4pt
\begin{picture}(3,2)(0,0.3)
\multiput(0,0)(1,0){2}{\line(0,1){2}}
\put(0,0){\line(1,0){1}}
\multiput(0,1)(0,1){2}{\line(1,0){3}}
\multiput(2,1)(1,0){2}{\line(0,1){1}}
\end{picture} .
Similarly, in the denominator, we join lower right corner of box ``1"
with all outer corners of 
\unitlength 4pt
\begin{picture}(3,2)(0,0.3)
\multiput(0,0)(1,0){2}{\line(0,1){2}}
\put(0,0){\line(1,0){1}}
\multiput(0,1)(0,1){2}{\line(1,0){3}}
\multiput(2,1)(1,0){2}{\line(0,1){1}}
\end{picture} .
For other factors, see the general formula given below.

\textbf{General formula}

Introduce the parameterization (\ref{para}) on $Y$.
Then the formula for $\langle A_{r+1,s+1}|J_Y L_{-n}$ is given as follows.

\textbf{Step 1:}
Remove box $(r_{i_1}^{(1)},s_{i_1}^{(1)})$ from $Y$ and denote it as
$Y\setminus (r_{i_1}^{(1)},s_{i_1}^{(1)})$.
Associated coefficient is
\begin{eqnarray}
  & & \frac{\displaystyle \prod_{j=0}^{i_1 -1}\left[ a(r_j^{(1)}
  +1,s_{i_1}^{(1)})+\left( l(r_j^{(1)}+1,s_{i_1}^{(1)})+1 \right)
  \beta\,\right] }{ \displaystyle \prod_{j=1}^{i_1 -1}\left[
  a(r_j^{(1)},s_{i_1}^{(1)})+l(r_j^{(1)},s_{i_1}^{(1)})\beta\, \right] }\\
  & & \times\frac{ \displaystyle \prod_{j={i_1}+1}^{m^{(1)}+1} \left[\left(
  a(r_{i_1}^{(1)},s_j^{(1)}+1)+1\right) +
  l(r_{i_1}^{(1)},s_j^{(1)}+1)\beta\, \right]}{ \displaystyle \prod_{j=i_1
  +1}^{m^{(1)}}\left[ a(r_{i_1}^{(1)},s_j^{(1)})+
  l(r_{i_1}^{(1)},s_j^{(1)})\beta\, \right] }.\nonumber
\end{eqnarray}

\textbf{Step 2:} 
Take a coordinate of $Y\setminus (r_{i_1}^{(1)},s_{i_1}^{(1)})$ as in $Y$,
and remove the box $(r_{i_2}^{(2)},s_{i_2}^{(2)})$. We obtain the same
coefficient as in Step 1, except that in the numerator, we don't incorporate
term corresponding to the corner created by the last removal.

\textbf{Step 3:}
Repeat Step 2 until $n$-th box is removed.

\textbf{Step 4:}
Multiply
\begin{equation}
  \frac{1}{n \beta^n}(-1)^{\#\{k|r_{i_k}^{(k)}>r_{i_{k+1}}^{(k+1)}\} }
  J_{Y\setminus \left( (r_{i_1}^{(1)},s_{i_1}^{(1)})\cup\cdots \cup
  (r_{i_n}^{(n)},s_{i_n}^{(n)}) \right)},
\end{equation}
to the result of Step 3.

\textbf{Step 5:}
Multiply 
\begin{equation}
  \langle A_{r+1,s+1}| \frac{1}{\sqrt{2\beta}} A_{\sum_{k=1}^{n} \left(
  r+3k-(r_{i_n}^{(n)}+r_{i_k}^{(k)}) \right), \sum_{k=1}^{n} \left(
  s+3k-(s_{i_n}^{(n)}+s_{i_k}^{(k)}) \right)},
\end{equation}
to the result of Step 4.

\textbf{Step 6:}
Repeat Steps 1 to 5 for each way to remove $n$ box from $Y$, and sum up all
the terms.

\begin{remark} 
  Formula for $\frac{\partial}{\partial p_n}J_Y$ is obtained by doing steps
  1 to 4 and 6.
\end{remark}

\begin{remark} 
  So far, we have considered leftward actions of $L_n$'s only. 
  To obtain formula $L_n J_Y |A\rangle$, we use the
  identity
  \begin{equation}
    \left( \langle J_{Y^{'}} | L_n\right) |J_Y\rangle = \langle J_{Y^{'}} |
    \left( L_n | J_Y\rangle\right),
  \end{equation}
  and orthogonality of Jack symmetric functions. Especially, in $L_n |J_Y
  \rangle$, the factor $A_{r-i,s-j}$ appears when the first removed box has
  coordinate $(i,j)$. We use this property in the next section.
\end{remark}

\begin{remark} 
  Although we did not give rigorous mathematical proof of the formulas
  presented in this section (except for the simplest non-trivial case $p_2J_Y$),
  numerous calculations involving Jack symmetric functions up to degree 10
  strongly support these results.
\end{remark}

\section{Application to representation theory}

\subsection{Singular vectors of Virasoro algebra}

We consider the Verma module defined by
\begin{equation}
  M(h)=\mathbf{C} [L_{-1},L_{-2},L_{-3},\cdots ]|h\rangle ,
\end{equation}
where $|h\rangle$ is a highest weight vector defined as $L_n |h\rangle =0$
for $n>0$ and $L_0 |h\rangle =h |h\rangle$. Then, from the Kac determinant
formula \cite{Kac,FF1,FF2}, we know that there is a singular vector of 
degree $rs$ if the highest weight is
\begin{equation}
  h_{r,s}=\frac{(r\beta -s)^2 -(\beta -1)^2}{4\beta}.
\end{equation}
We denote this singular vector as $|\chi _{r,s}\rangle $.

To compare the Verma module and the Fock space, using the bosonic 
representation (\ref{FeFu}), we have
\begin{equation}
  L_0 |A_{r+1,s+1}\rangle =
h_{r,s} |A_{r+1,s+1}\rangle ,
\end{equation}
since $h_{r,s}=\frac{1}{2}A_{r+1,s+1}A_{r-1,s-1}$.
When we bosonize $|\chi
_{r,s}\rangle $, Mimachi-Yamada's formula states that this is proportional to
the bosonized Jack symmetric function with the partition $(s^r)$, i.e. the
rectangle with horizontal length $s$ and vertical length $r$.

Our argument in the previous section gives us an intuitive reinterpretation of
this fact. Consider the Jack symmetric function with rectangular
partition of $m$ rows and $n$ columns and act by $L_1$ and $L_2$ respectively.
Then, ignoring the factors independent of integers $r$, $s$ in
$h_{r,s}$ for simplicity, we have
\begin{eqnarray}
L_1 J
\unitlength 3pt
\begin{picture}(4.5,3)(0,2)
\multiput(0,0)(4,0){2}{\line(0,1){3}}
\multiput(0,0)(0,3){2}{\line(1,0){4}}
\end{picture}
|A_{r+1,s+1}\rangle
&=&
A_{r-m,s-n} J
\unitlength 3pt
\begin{picture}(4.5,3)(0,2)
\put(0,0){\line(1,0){3}}
\put(0,0){\line(0,1){3}}
\put(0,3){\line(1,0){4}}
\put(3,0){\line(0,1){1}}
\put(3,1){\line(1,0){1}}
\put(4,1){\line(0,1){2}}
\end{picture}
|A_{r+1,s+1}\rangle ,\\
L_2 J
\unitlength 3pt
\begin{picture}(4.5,3)(0,2)
\multiput(0,0)(4,0){2}{\line(0,1){3}}
\multiput(0,0)(0,3){2}{\line(1,0){4}}
\end{picture}
|A_{r+1,s+1}\rangle
&=&
A_{r-m,s-n}\left(
J \unitlength 3pt
\begin{picture}(4.5,3)(0,2)
\put(0,0){\line(1,0){2}}
\put(0,0){\line(0,1){3}}
\put(2,0){\line(0,1){1}}
\put(2,1){\line(1,0){2}}
\put(4,1){\line(0,1){2}}
\put(0,3){\line(1,0){4}}
\end{picture}
+ J \unitlength 3pt
\begin{picture}(4.5,3)(0,2)
\put(0,0){\line(1,0){3}}
\put(0,0){\line(0,1){3}}
\put(3,0){\line(0,1){2}}
\put(3,2){\line(1,0){1}}
\put(4,2){\line(0,1){1}}
\put(0,3){\line(1,0){4}}
\end{picture}
\right)
|A_{r+1,s+1}\rangle .
\end{eqnarray} 
{}From the identity $A_{0,0}=0$, if we choose highest weight as $h_{m,n}$,
then above two actions vanish. Then from the commutation relation of the Virasoro
algebra, we have $L_p J \unitlength 3pt \begin{picture}(4.5,3)(0,2)
\multiput(0,0)(4,0){2}{\line(0,1){3}} \multiput(0,0)(0,3){2}{\line(1,0){4}}
\end{picture}
|A_{m+1,n+1}\rangle =0$ for general $p>0$.
Thus we can conclude that $J_{(n^m)}|A_{m+1,n+1}\rangle$ is the bosonized
singular vector of highest weight $h_{m,n}$.

\subsection{Refinement of Mimachi-Yamada formula}

In this section, we compute the singular vectors including the
proportionality factor also. Let us normalize the singular vector $|\chi
_{r,s}\rangle $ in such a way that the coefficient of $L_{-1}^{rs} $ is
equal to 1 \cite{Shi}. For example, we have
\begin{eqnarray}
  |\chi _{2,2} \rangle &=& \left( -\frac{3(1-\beta)^2}{\beta}L_{-4}
  -\frac{2(1-3\beta +\beta^2)}{\beta}L_{-3}L_{-1}\right. \nonumber \\
  & & \left. +\frac{(1-\beta^2)^2}{\beta^2}L_{-2}^2 -\frac{2(1+\beta
  ^2)}{\beta} L_{-2}L_{-1}^2 +L_{-1}^4 \right) |h_{2,2}\rangle .\nonumber
\end{eqnarray}
We calculate the bosonization of $|\chi _{r,s}\rangle $. However, since we
already know that this is proportional to the Jack symmetric function
$J_{(s^r )}$, the only thing we have to do is to determine the
proportionality factor. In this context, our normalization of Jack
symmetric function is useful, because the coefficient of $p_1^n$ is equal
to 1 for any Jack symmetric function of degree $n$. Thus, from
identification (\ref{liden}), we need to consider only the coefficient of
$a_{-1}^n$ in the bosonized singular vector of degree $n$.

First note that in this problem, we have a nice property.
We give two examples;
\begin{eqnarray}
  L_{-3}L_{-1}|A_{r+1,s+1}\rangle &=& \left(
  A_{r+1,s+1}a_{-4}+A_{r+3,s+3}A_{r+1,s+1}a_{-3}a_{-1}\right. \label{31}\\
  & &\left. +A_{r+1,s+1}a_{-2}a_{-1}^2 \right) |A_{r+1,s+1}\rangle ,
  \nonumber\\
  L_{-2}L_{-1}^2 |A_{r+1,s+1}\rangle &=& \left( 2A_{r+1,s+1}a_{-4}+\cdots
  +\frac{A_{r+1,s+1}^2}{2} a_{-1}^4 \right) |A_{r+1,s+1}\rangle
  .\label{211}
\end{eqnarray}
Comparing these two examples, the term $a_{-1}^4$ appears only in (\ref{211}),
and it does not appear in (\ref{31}). In general, we can show that terms like
$a_{-1}^n$ appear only in the product of $L_{-1}$ and $L_{-2}$, and their
coefficients have simple forms.

For this purpose, we define the subset $H \subset \mathcal{F}_A$ as
\begin{equation}
  H=\mathrm{span}\Bigl\{ a_{-1}^{i_1}a_{-2}^{i_2}\cdots a_{-n}^{i_n} |A\rangle
  \Bigl| i_k \neq 0\,\, \mathrm{for\,\, some}\,\,k>1 \Bigl\} .
\end{equation}
For example, $a_{-2}a_{-1}^2\in H$ and $a_{-1}^4\notin H$.
Then we have

\textbf{Lemma} \textit{For some $g\in H$, we have}
\begin{equation}
  L_{-n}^{i_n}\cdots L_{-2}^{i_2} L_{-1}^{i_1} |A\rangle \, = 
  \left\{
  \begin{array}{ll}
    {\frac{A^{i_1}}{2^{i_2}}}a_{-1}^{i_1 +2i_2}|A\rangle+g &
    (i_3=i_4=\cdots =i_n=0) \\
    g  & (\mathrm{otherwise}) \\
  \end{array}
  \right..
\end{equation}

\textbf{Proof} is by induction.

\textbf{Step 1:} We show 
\begin{equation}
  L_{-1}^{i_1}|A\rangle = A^{i_1}a_{-1}^{i_1}|A\rangle +g\,\, (g\in H).
\label{step1}
\end{equation}
{}From representation (\ref{FeFu}), $L_{-1}$ is expressed as a sum of terms
$a_{-i-1}a_i$ $(i\geq 0)$. {}From the condition $a_n |A\rangle =0$ $(n>0)$, we
have $L_{-1}|A\rangle =A a_{-1}|A\rangle$. Let us assume for some $n\in \mathbf{Z}_{\geq 0}$, the
relation $L_{-1}^n |A\rangle = A^n a_{-1}^n| A\rangle +g_1$ $(g_1\in H)$
holds. We can easily see that
\begin{equation}
  a_{-i-1}a_i \cdot a_{-m}^{l_m}a_{-(m-1)}^{l_{m-1}}\cdots
  a_{-1}^{l_1}|A\rangle = a_{-m}^{l_m}\cdots a_{-(i+1)}^{l_{i+1}+1}\,
  il_i\, a_{-i}^{l_i-1}\cdots a_{-1}^{l_1} |A\rangle \,\,\in H ,
\end{equation}
if $m\geq i$ (with a slight modification if $m=i$), otherwise act as 0.
This means 
\begin{equation}
  \sum_{i=1}^{\infty}a_{-i-1}a_i \cdot L_{-1}^n |A\rangle\in H.
\end{equation}
We also notice that for $g\in H$, $a_{-1}a_0 \cdot g\in H$.
Therefore we have, for some $g_2\in H$
\begin{eqnarray}
  L_{-1}\cdot L_{-1}^n |A\rangle &=& a_{-1}a_0(A^n a_{-1}^n |A\rangle
  )+g_2\nonumber \\
  &=& A^{n+1}a_{-1}^{n+1}|A\rangle +g_2 .
\end{eqnarray}
{}From induction hypothesis, we conclude that (\ref{step1}) 
is valid for any $i_1\in \mathbf{Z}_{\geq 0}$.

\textbf{Step 2:}
We consider the action of $L_{-2}$, whose representation is given by
\begin{equation}
  L_{-2}=A_{1,1}a_{-2}+\frac{1}{2}a_{-1}^2+a_{-2}a_0+a_{-3}a_1+a_{-4}a_2 +
  \cdots .
\end{equation}
{}From the definition of $H$, we notice that $A_{1,1}a_{-2} \cdot
\mathcal{F}_A \subset H$. On the other hand, terms like $a_{-i-2}a_i$ $(i>0)$ act as
\begin{equation}
  a_{-i-2}a_i \cdot a_{-m}^{l_m}a_{-(m-1)}^{l_{m-1}}\cdots
  a_{-1}^{l_1}|A\rangle = a_{-m}^{l_m}\cdots
  a_{-(i+2)}^{l_{i+2}+1}a_{-(i+1)}^{l_{i+1}}\, il_i\, a_{-i}^{l_i -1}\cdots
  a_{-1}^{l_1}|A\rangle\,\,\in H.
\end{equation}
Thus, as in Step 1, we have
\begin{equation}
  L_{-2}^{i_2}L_{-1}^{i_1}|A\rangle =
\frac{A^{i_1}}{2^{i_2}}a_{-1}^{i_1 +2i_2}|A\rangle +g\,\,(g\in H)
\end{equation}
for general $i_1,i_2\in \mathbf{Z}_{\geq 0}$.

\textbf{Step 3:} We consider $L _{-n}^{i_n}\cdots L_{-2}^{i_2} L_{-1}^{i_1}
|A\rangle$ in case that for some $n>2$, $i_n\neq 0$. By the representation
(\ref{FeFu}), this $L_{-n}$ is a sum of terms $a_{-n}$ and $a_{-i}a_{-j}$
$(i+j=n)$ with some coefficients. By the condition $n>2$, we have $a_{-n}\cdot
\mathcal{F}_A\subset H$. We also have $\mathrm{max}(i,j)\geq
\frac{n}{2}>1$, in each $a_{-i}a_{-j}$ $(i+j=n)$, there is at least one
operator $a_{-k}$, $k>1$. Therefore $a_{-i}a_{-j}\cdot \mathcal{F}_A
\subset H$, if $(i+j=n)$, and we have
\begin{equation}
  L _{-n}^{i_n}\cdots L_{-2}^{i_2} L_{-1}^{i_1} |A\rangle \,\in H\quad (n>2).
\end{equation}
The proof of Lemma is now finished.
\rule{5pt}{10pt}

Thus, to calculate the proportionality factor we are interested in, we need
only information about terms like $L_{-2}^{i_2}L_{-1}^{i_1}|h\rangle$ in
the singular vector, and this has already been calculated by Feigin-Fuchs
\cite{FF2} (with reference to \cite{L}). To quote their result, we prepare
some notations (see for example \cite{BDIZ} for detail). For each positive
integer $k$, consider the subalgebra generated by $L_{-k}$, $L_{-k-1}$,
$L_{-k-2}$, $\cdots$, and denote its enveloping algebra by $U_{-k}$. Let
us write the singular vectors on the quotient algebra $U_{-1}/U_{-3}$ as
\begin{equation}
  |\chi _{r,s}\rangle =\sigma_{j^{'}j}(L_{-1},L_{-2})|h\rangle ,
\end{equation}
where $r=2j^{'}+1$, $s=2j+1$.
Then we have
\begin{equation}
  \sigma_{j^{'}j}(L_{-1},L_{-2})^2=\prod_{\scriptstyle -j\leq M\leq j \atop
  \scriptstyle -j^{'}\leq M^{'}\leq j^{'} } \left(L_{-1}^2 +4(M\theta
  +M^{'}\theta^{-1})^2 L_{-2}\right) ,
  \label{LFF}
\end{equation}
where $\theta^2 =-1/\beta$.
Note that we have $[L_{-1},L_{-2} ]=0$ on $U_{-1}/U_{-3}$.

Then, if we bosonize the factor $\left(L_{-1}^2 +4(M\theta
+M^{'}\theta^{-1})^2 L_{-2} |h_{r,s}\rangle\right)$ and use above Lemma and
identification (\ref{liden}), we obtain the coefficient of $p_1^2$ as
\begin{eqnarray}
  & & \left(A_{r+1,s+1}^2 +2\left(-\frac{M^2}{\beta}+2MM^{'}-{M^{'}}^2
  \beta \right)\right)\frac{\beta}{2} \nonumber \\
  & &= [(j^{'}+1+M^{'})\beta -(j+1+M)]\times[(j^{'}+1-M^{'})\beta -(j+1-M)].
\end{eqnarray}
Take product of above expression as in (\ref{LFF}), and convert the result
in language of symmetric functions. Then we notice that in the image of
$\sigma_{j^{'}j}(L_{-1},L_{-2})^2 |h_{r,s}\rangle$, the coefficient of
$p_1^{rs}$ is given by $\prod_{i=1}^r \prod_{j=1}^s (i\beta -j)$. 
Since the coefficient of $p_1^{rs}$ in the integral form of 
the Jack symmetric function of
degree $rs$ is always 1, this factor itself is equal to the proportionality factor
between the singular vectors in the Verma module and the Fock space. Thus,
we have the following formula.

\textbf{Formula}
\begin{equation}
  |\chi _{r,s}\rangle = \prod_{i=1}^r \prod_{j=1}^s (i\beta -j) J_{(s^r)}
  |A_{r,s}\rangle
\end{equation}

In the same way, we can also show that
\begin{equation}
  \langle\chi _{r,s}| =0 .
\end{equation}
To verify this, we first observe following two equations,
\begin{eqnarray}
  L_{-n}|A_{r+1,s+1}\rangle &=&
  [(n-1)A_{1,1}a_{-n}+A_{r+1,s+1}a_{-n}+a_{-n+1}a_{-1}+\cdots
  ]|A_{r+1,s+1}\rangle ,\label{left}\\
  \langle A_{r+1,s+1} |L_n &=& \langle A_{r+1,s+1} |[-(n+1)A_{1,1}a_n
  +A_{r+1,s+1} a_n +a_{n-1} a_1 +\cdots ].
\end{eqnarray}
In the above equations, the expansion is in fact a finite sum.  Comparing these
two equations, we notice that to obtain $\langle A_{r+1,s+1} |L_n$ from the
result of $L_{-n}|A_{r+1,s+1}\rangle$, we just replace
$(A_{r+1,s+1},A_{1,1},a_{-i})$ in (\ref{left}) by 
\begin{equation}
(A_{r+1,s+1},A_{1,1},a_{-i})\longrightarrow (A_{r-1,s-1},A_{-1,-1},a_{i}).
\label{replacement}
\end{equation}
 The same relation holds if we consider a general vector
\begin{equation}
  L_{-n}^{i_n}\cdots L_{-2}^{i_2} L_{-1}^{i_1} |A_{r+1,s+1}\rangle
\end{equation} 
and its dual
\begin{equation}
  \langle A_{r+1,s+1}| L_1^{i_1} L_2^{i_2}\cdots L_n^{i_n},
\end{equation} since $[a_n,a_0]=0\, (n\in \mathbf{Z})$.
Thus we can easily obtain dual of above Lemma.

Since $|\chi_{r,s}\rangle$ is proportional to $J_{(s^r)}$
in terms of $p_n =\sqrt{\frac{2}{\beta}}a_{-n}$, we can use the replacement (\ref{replacement}).
Then we notice that $\langle\chi_{r,s}|$ has the same structure as $J_{(s^r)}$,
if we use $\tilde{p}_n:=2p_n$ in stead of $p_n$
(compare the two identifications (\ref{liden}) and (\ref{riden})).
Thus, in this case, we also have to consider the coefficient in front of $p_1^{rs}$.
To do this, we consider the dual of (\ref{LFF}). In particular, we have to
calculate the term $(M,M^{'})=(j,j^{'})$, i.e.
\begin{equation}
  \left( L_1^2 +4(j\theta +j^{'} \theta^{-1})^2L_2 \right) .
\end{equation}
Then from the dual of Lemma and identification (\ref{riden}), we have
\begin{equation}
  \left(A_{r-1,s-1}^2+2(j\theta +j^{'}\theta^{-1})^2\right) 2\beta =0.
\end{equation}
Thus we obtain $\langle \chi_{r,s}|=0$ as stated.

\begin{remark} 
  Result in this section can be viewed as determination of
  proportionality factor 
  between bosonized Jack symmetric functions and its analogue in the Verma
  module. Thus we can expect that similar coefficient occurs when general
  Jack symmetric function is embedded in Verma module.
\end{remark}

\begin{remark}
After completion of the manuscript, we noticed a paper of Adler and
van Moerbeke \cite{AvM}. They calculate the action of Virasoro operators
on Schur symmetric functions.
We also noticed a paper by B. Feigin and E. Feigin \cite{FeFe}.
They consider the bosonized Jack symmetric functions as a basis of the Fock
space and studied the representation of vertex operator algebra.
\end{remark}

\textbf{Acknowledgments:} One of the author (R.S.) is grateful to Miki
Wadati for continuous encouragement. 
R.S. and J.S. are grateful to Boris Feigin, Omar Foda and Michio Jimbo
for discussion. 
The authors would like to thank CNRS 
for financial support under PICS contract number 911.


\end{document}